\documentclass[12pt]{iopart}

\usepackage{graphicx}
\usepackage{epsfig}
\usepackage{amsfonts}
\usepackage{amssymb}

\newcommand{\vphi}{\varphi}

\begin{document}

\title[Rotating black holes with non-Abelian hair]{Rotating black holes with non-Abelian hair}

\author{Burkhard Kleihaus$^1$, Jutta Kunz$^1$ and
Francisco Navarro-L\'erida$^2$}

\address{$^1$ Institut f\"ur Physik, Universit\"at Oldenburg, 26111 Oldenburg, Germany}
\address{$^2$ Dept.~de F\'{\i}sica At\'omica, Molecular y Nuclear, Ciencias F\'{\i}sicas\\
Universidad Complutense de Madrid, E-28040 Madrid, Spain}
\ead{b.kleihaus@uni-oldenburg.de, jutta.kunz@uni-oldenburg.de,
fnavarro@fis.ucm.es}
\vspace{10pt}
\begin{indented}
\item[]March 2016
\end{indented}

\begin{abstract}
We here review asymptotically flat rotating black holes in the presence of non-Abelian gauge fields.
Like their static counterparts these black holes are no longer uniquely determined by their
global charges. In the case of pure SU(2) Yang-Mills fields, the rotation 
generically induces an electric charge,
while the black holes do not carry a magnetic charge. When a Higgs field is coupled, 
rotating black holes with monopole hair arise in the case of a Higgs triplet, while
in the presence of a complex Higgs doublet the black holes carry sphaleron hair.
The inclusion of a dilaton allows for Smarr type mass formulae.
\end{abstract}

%
\vspace{2pc}

\normalsize

\section{Introduction}
Asymptotically flat electrovac black holes in four spacetime dimensions
are remarkably simple objects.
They are comprised in the Kerr-Newman (KN) family, and uniquely characterized 
by their global charges: the mass, the angular momentum and the
electromagnetic charges 
\cite{Israel:1967za,Robinson:1975bv,Mazur:1982db,Chrusciel:2012jk}.
Their simplicity and beauty led to the conjecture, that the properties
of this electrovac family of black holes were generic properties of black holes
and that the theorems holding for them should remain true 
in the presence of other fields, as well. In particular, 
black holes should not carry hair and thus not possess any special features
beyond those that would be fully determined by their global charges.

Towards the end of the last century,
the consideration of non-Abelian fields coupled to Einstein gravity
had a major impact on our understanding of the properties of black holes
in four dimensions. The first hairy black holes constructed at the time
were black holes with a particular set of non-Abelian scalar fields, i.e.,
black holes carrying Skyrme hair
\cite{Luckock:1986tr,Droz:1991cx}.
These static hairy black holes emerge from gravitating Skyrmion solutions,
when the presence of a small horizon is imposed in the interior of the
solitonic configurations. However, the horizon size of these
hairy black holes is bounded from above.
So large static black holes in Einstein-Skyrme theory can only be Schwarzschild
black holes.

Soon thereafter, the discovery of the 
globally regular Bartnik-McKinnon solutions \cite{Bartnik:1988am}
in SU(2) Einstein-Yang-Mills (EYM) theory led to the quest for 
black holes with Yang-Mills hair 
\cite{Volkov:1990sva,Bizon:1990sr,Kuenzle:1990is}.
The resulting static spherically symmetric EYM black holes possess 
a purely magnetic gauge field and a single
global charge, their mass. For a given horizon radius
there is an infinite number of black holes, all differing in their metric and
field configurations. In particular, these
black holes can be labelled by an integer $k$, counting the number of nodes of
the single gauge field function. 
Obviously, the global charges no longer determine these non-Abelian
black holes uniquely.
While there is no limit on the mass and the
size of these black holes, with increasing integer $k$ these
magnetically neutral solutions tend towards
some limiting configuration, which 
in its outer part corresponds to an extremal 
Reissner-Nordstr\"om (RN) black hole with unit magnetic charge
\cite{Smoller:1994ks,Breitenlohner:1993es,Breitenlohner:1994qc}.

However, besides their infinite non-uniqueness, these EYM black holes
held further surprizes. For electrovac black holes Israel's theorem states
that static black holes are spherically symmetric 
\cite{Israel:1967za}.
In EYM theory this is no longer true. Indeed,
static EYM black holes exist, which possess only axial symmetry 
\cite{Kleihaus:1997ic,Kleihaus:1997ws}.
When the circumference of the horizon is measured along the poles
and compared to the circumference measured along the equator,
their ratio differs from one.
Such deformed static black holes can be obtained by including
a winding number $n$ w.r.t.~the azimuthal angle, leading
to prolate horizon shapes. By including an integer w.r.t.~the polar angle,
one obtains a further type of static EYM black holes 
\cite{Ibadov:2005rb}.
All these configurations can be interpreted as bound states
of black holes and solitonic configurations 
\cite{Ashtekar:2000hw,Corichi:1999nw,Corichi:2000dm,Ashtekar:2000nx,Kleihaus:2001ti}.

In this review we will first focus on rotating EYM black hole solutions 
\cite{Volkov:1997qb,Brodbeck:1997ek,Kleihaus:2000kg,Kleihaus:2002ee}
(for a review on the static case see e.g.~\cite{Volkov:1998cc}).
These black holes represent generalizations of the
KN solutions, since they carry mass, angular momentum and a small
electric charge, which is induced by the rotation 
\cite{Volkov:1997qb}.
It is interesting to note, that whereas the static 
axially symmetric EYM black holes possess globally regular counterparts
\cite{Kleihaus:1996vi,Kleihaus:1997mn,Ibadov:2004rt},
the rotating EYM black holes do not
\cite{VanderBij:2001nm,vanderBij:2002sq}.
When trying to obtain such solitonic rotating configurations
with a regular center the fall-off conditions at spatial infinity cannot be met 
when only massless fields are present.
The EYM black holes can be generalized by including a dilaton field.
In its presence a Smarr type mass formula
holds \cite{Kleihaus:2002tc}. Moreover, such rotating 
Einstein-Yang-Mills-dilaton (EYMd) black holes need not carry
an electric charge in the presence of rotation \cite{Kleihaus:2003sh}.

Subsequently, we will 
discuss the solutions obtained by including a Higgs field, giving
mass to (some of) the gauge fields. In the case of a Higgs triplet,
gravitating monopole solutions \cite{Lee:1991vy,Breitenlohner:1991aa,Breitenlohner:1994di,Brihaye:1998cm},
monopole-antimonopole pairs and systems arise 
\cite{Kleihaus:2000hx,Hartmann:2000gx,Hartmann:2001ic,Kleihaus:2004fh,Kunz:2007jw},
which can be endowed with a black hole at their center 
\cite{Lee:1991vy,Breitenlohner:1991aa,Breitenlohner:1994di,Hartmann:2000gx,Hartmann:2001ic,Kleihaus:2004fh,Kleihaus:2005fs,Kleihaus:2007vf,Kleihaus:2000kv}.
While all of these black holes can be put into rotation, by giving their
horizon a finite angular velocity \cite{Kleihaus:2007vf,Kleihaus:2004gm}, 
only the globally regular solutions
without a global magnetic charge can carry angular momentum
\cite{VanderBij:2001nm,vanderBij:2002sq,Kleihaus:2005fs,Paturyan:2004ps}.
Interestingly, the rotating hairy black holes bifurcate with KN black holes.
Thus KN black holes are unstable w.r.t.~growing Yang-Mills-Higgs (YMH) hair.
In the presence of a dilaton again a Smarr type mass formula
holds \cite{Kleihaus:2007vf}.

From a physical point of view the Einstein-Yang-Mills-Higgs (EYMH) black holes
with a Higgs triplet might be relevant in the early universe around a
Grand Unified Theory (GUT) phase transition,
since a YMH theory with a Higgs triplet may be embedded in 
relevant GUTs. Later, during the electroweak phase transition,
the Standard Model (SM) gauge and Higgs fields play a prominent role.
Here a complex doublet of Higgs fields is needed to give the
electroweak vector bosons mass, while the photon remains massless.
The globally regular solutions of this theory are electroweak sphalerons
and their generalizations \cite{Klinkhamer:1984di,Boguta:1983xs,Kleihaus:1991ks}.
Coupling them to
Einstein gravity and endowing them with a black hole at their center,
leads again to bound states of static black holes
and globally regular solutions 
\cite{Greene:1992fw,Brihaye:2000ez,Ibadov:2008hj}.
We will address their rotating generalizations towards the end
of this review \cite{Radu:2008ta,Kleihaus:2008cv,Ibadov:2010ei}.

Whereas all of these rotating hairy black hole solutions involve
non-Abelian fields, recently also rotating hair black holes of a different kind
were found \cite{Herdeiro:2014goa}. These are based on the presence of a 
single complex scalar field
and thus a U(1) invariance of the theory. Unlike the non-Abelian
black holes addressed in this review, they possess no non-rotating limit.
But they may be interpreted as bound states of black holes and solitons, as well,
representing black holes inside rotating boson stars.
Many of their properties have been studied by now in detail
including their ergosurfaces, their shadows or their iron-K$\alpha$ lines
\cite{Herdeiro:2014ima,Herdeiro:2014jaa,Herdeiro:2015gia,Cunha:2015yba,Herdeiro:2015tia,Herdeiro:2016gxs,Vincent:2016sjq,Ni:2016rhz,Delgado:2016jxq,Cunha:2016bjh}.
The explicit time-dependence of the complex scalar field
seems to make them generically different from the EYM and EYMH black holes.
However, for the latter an analogous time-dependence
signalling a rotation in internal space will arise in a different gauge,
when the black holes carry electric charge.
Likewise, if one were to rotate black holes with Skyrmion hair, 
the scalar field would need an explicit time dependence
\cite{Ioannidou:2006nn}.
Note also the recent review article on black holes with scalar hair
\cite{Herdeiro:2015waa}.

In section 2 of this review
rotating black holes with YM and YM dilaton (YMD) hair are addressed.
In section 3 a Higgs triplet is included, and the new features of the
resulting rotating black holes are discussed.
For a Higgs doublet only static black holes
have been studied so far.
We conclude in section 4.

\section{Rotating Einstein-Yang-Mills-dilaton black holes}

Here we present the current status of asymptotically flat rotating
Einstein-Yang-Mills (EYM) and Einstein-Yang-Mills-dilaton (EYMD)
black holes. Starting with the action and field equations for the general case,
we then address the physical properties and exhibit a number of interesting
results for these rotating non-Abelian black holes, first for the case of
EYM black holes obtained for vanishing dilaton coupling constant $\gamma$,
and subsequently for EYMD black holes with arbitrary $\gamma$.

\subsection{Action}

The SU(2) 
EYMD action is given by
\begin{equation}
S=\int \left ( \frac{R}{16\pi G} + L_M \right ) \sqrt{-g} d^4x \ ,
\ \label{action} \end{equation}
with scalar curvature $R$ and matter Lagrangian $L_M$,
\begin{equation}
L_M=-\frac{1}{2}\partial_\mu \Psi \partial^\mu \Psi
 - \frac{1}{2} e^{2 \gamma \Psi } {\rm Tr} (F_{\mu\nu} F^{\mu\nu})
\ , \label{lagm} \end{equation}
with dilaton field $\Psi$,
field strength tensor $ F_{\mu \nu} = 
\partial_\mu A_\nu -\partial_\nu A_\mu + i e \left[A_\mu , A_\nu \right] $,
gauge field $A_\mu =  A_\mu^a \tau_a/2$,
and Newton's constant $G$, dilaton coupling constant $\gamma$,
and Yang-Mills coupling constant $e$.

Variation of the action with respect to the metric and the matter fields
leads to the Einstein equations 
\begin{equation}
G_{\mu\nu}= R_{\mu\nu}-\frac{1}{2}g_{\mu\nu}R = 8\pi G T_{\mu\nu}
\ ,  \label{ee} \end{equation}
with stress-energy tensor
\begin{eqnarray}
T_{\mu\nu} &=& g_{\mu\nu}L_M -2 \frac{\partial L_M}{\partial g^{\mu\nu}}
 \nonumber \\
  &=& \partial_\mu \Psi \partial_\nu \Psi
     -\frac{1}{2} g_{\mu\nu} \partial_\alpha \Psi \partial^\alpha \Psi
      + 2 e^{2 \gamma \Psi }{\rm Tr}
    ( F_{\mu\alpha} F_{\nu\beta} g^{\alpha\beta}
   -\frac{1}{4} g_{\mu\nu} F_{\alpha\beta} F^{\alpha\beta})
\ , \label{tmunu}
\end{eqnarray}
and the matter field equations,
\begin{eqnarray}
& &\frac{1}{\sqrt{-g}} \partial_\mu \left(\sqrt{-g} 
 \partial^\mu \Psi \right)=
 \gamma e^{2 \gamma \Psi } 
  {\rm Tr} \left( F_{\mu\nu} F^{\mu\nu} \right)  \ ,
\label{feqD} \end{eqnarray}
\begin{eqnarray}
& &\frac{1}{\sqrt{-g}} D_\mu(\sqrt{-g} e^{2 \gamma \Psi } F^{\mu\nu}) = 0 \ ,
\label{feqA} \end{eqnarray}
where $D_{\mu}=\partial_{\mu} + ie \left[A_\mu , \cdot \right]$.

\subsection{Symmetries and Ans\"atze}

Considering black hole solutions,
which are both stationary and axially symmetric,
we impose on the spacetime the presence of
two commuting Killing vector fields, 
$\xi$ (asymptotically timelike) and $\eta$ (asymptotically spacelike),
and adopt a system of adapted coordinates, 
$\{t, r, \theta, \varphi\}$, such that 
\begin{equation}
\xi=\partial_t \ , \ \ \ \eta=\partial_{\varphi}
\ . \label{xieta} \end{equation}
In these coordinates the metric is independent of $t$ and $\varphi$,

In addition to the symmetry requirements on the metric 
(${\cal L}_{\xi} g = {\cal L}_{\eta} g =0$),
we require that the matter fields satisfy the same symmetries.
For dilaton this implies
\begin{equation}
{\cal L}_{\xi} \Psi = {\cal L}_{\eta} \Psi =0
\ ,  \label{symd} \end{equation}
whereas for the non-Abelian gauge potential $A=A_\mu dx^\mu$ 
this implies (because of the non-Abelian gauge symmetry)
\cite{Forgacs:1979zs}
\begin{eqnarray}
\displaystyle ({\cal L}_{\xi}A)_{\mu} &=&  D_{\mu} W_{\xi}
\ , \nonumber \\
\displaystyle ({\cal L}_{\eta}A)_{\mu} &=&  D_{\mu} W_{\eta}
\ , \label{symA} \end{eqnarray}
where $W_{\xi}$ and $W_{\eta}$ are su(2)-valued functions 
which satisfy
\begin{equation}
{\cal L}_{\xi}W_{\eta}-{\cal L}_{\eta}W_{\xi} 
                + i e \left[W_{\xi},W_{\eta}\right] = 0
\ .  \label{compat} \end{equation}
Exploiting the gauge invariance to set $W_{\xi}=0$
leaves $A$ and $W_{\eta}$ independent of $t$.

The metric can then be written in the Lewis-Papapetrou 
form, which in isotropic coordinates reads
\begin{equation}
ds^2 = -fdt^2+\frac{h}{f}\left[dr^2+r^2 d\theta^2\right] 
       +\sin^2\theta r^2 \frac{l}{f}
          \left[d\vphi-\frac{\omega}{r}dt\right]^2 \  
\ , \label{metric} \end{equation}
and the functions $f$, $h$, $l$ and $\omega$ depend on $r$ and $\theta$, only.
Regularity along the symmetry axis then requires
\begin{equation}
h|_{\theta=0,\pi}=l|_{\theta=0,\pi}
\ . \label{lm} \end{equation}

The event horizon resides at a surface of constant radial coordinate, $r=r_{\rm H}$,
and is characterized by the condition $f(r_{\rm H},\theta)=0$ \cite{Kleihaus:2000kg}.
The Killing vector field
\begin{equation}
\chi = \xi + \frac{\omega_{\rm H}}{r_{\rm H}} \eta = \xi + \Omega \eta
\ , \label{chi} \end{equation}
is orthogonal to and null on the horizon, and $\Omega$ is the horizon angular
velocity.

The ergoregion is defined as the region outside the event horizon
where $\xi_\mu \xi^\mu$ is positive.
It is bounded by the ergosurface where
\begin{equation}
 -f +\sin^2\theta \frac{l}{f} \omega^2 = 0 \ .
 \label{ergo}
\end{equation}

For the gauge fields one can employ a rather general ansatz \cite{Kleihaus:2007vf}, 
fulfilling the symmetry constraints 
\begin{eqnarray}
A_\mu dx^\mu
 & = &  \left( B_1 \frac{\tau_r^{(n,m)}}{2e} + B_2 \frac{\tau_\theta^{(n,m)}}{2e}
 \right) dt 
+A_\varphi (d\varphi-\frac{\omega}{r} dt)\nonumber \\
& + &\left(\frac{H_1}{r}dr +(1-H_2)d\theta \right)\frac{\tau_\varphi^{(n)}}{2e}
 , \label{a1} \end{eqnarray}
\begin{equation}
A_\varphi=   -n\sin\theta\left(H_3 \frac{\tau_r^{(n,m)}}{2e}
            +(1-H_4) \frac{\tau_\theta^{(n,m)}}{2e}\right)  
 , \label{a2} \end{equation}
where $n$ and $m$ are integers,
and the symbols $\tau_r^{(n,m)}$, $\tau_\theta^{(n,m)}$ and $\tau_\vphi^{(n)}$
denote the dot products of the Cartesian vector of Pauli matrices,
$\vec \tau = ( \tau_x, \tau_y, \tau_z) $,
with the spatial unit vectors
\begin{eqnarray}
{\hat e}_r^{(n,m)} & = & \left(
\sin(m\theta) \cos(n\vphi), \sin(m\theta)\sin(n\vphi), \cos(m\theta)
\right)\ , \nonumber \\
{\hat e}_\theta^{(n,m)} & = & \left(
\cos(m\theta) \cos(n\vphi), \cos(m\theta)\sin(n\vphi), -\sin(m\theta)
\right)\ , \nonumber \\
{\hat e}_\vphi^{(n)} & = & \left( -\sin(n\vphi), \cos(n\vphi), 0 \right)\ ,
\label{unit_e}
\end{eqnarray}
respectively.
Like the dilaton field function $\Psi$,
the gauge field functions $B_i$ and $H_i$
depend only on the coordinates $r$ and $\theta$.

We  note that the ansatz is form-invariant under gauge transformations $U$
\cite{Kleihaus:2000kg,Kleihaus:2007vf}
\begin{equation}
 U= \exp \left({\frac{i}{2} \tau^{(n)}_\varphi \Gamma(r,\theta)} \right)
\ .\label{gauge} \end{equation}
With respect to this residual gauge degree of freedom 
one can choose the gauge fixing condition 
$r\partial_r H_1-\partial_\theta H_2 =0$.
%
We also remark that the above Ansatz satisfies the circularity conditions.

\subsection{Boundary conditions}

\noindent {\sl Dimensionless Quantities:} It is convenient 
to introduce the dimensionless coordinate $\bar r$,
\begin{equation}
\bar r=\frac{e}{\sqrt{4\pi G}} r
\ , \label{dimless} \end{equation}
the dimensionless electric gauge field functions
${\bar B}_1$ and ${\bar B}_2$,
\begin{equation}
{\bar B}_1 = \frac{\sqrt{4 \pi G}}{e}  B_1 \ , \ \ \
{\bar B}_2 = \frac{\sqrt{4 \pi G}}{e}  B_2 \ ,
\label{barb} \end{equation}
the dimensionless dilaton function $\bar \Psi$,
\begin{equation}
\bar \Psi = \sqrt{4\pi G} \Psi
\ , \label{dimp} \end{equation}
and the dimensionless dilaton coupling constant $\bar \gamma$,
\begin{equation}
\bar \gamma =\frac{1}{\sqrt{4\pi G}} \gamma
\ , \label{dimg} \end{equation}
and subsequently omit the bar for notational simplicity.
For $\gamma = 0$
the dilaton decouples and EYM theory is obtained.

\noindent {\sl Boundary conditions at infinity:} To obtain asymptotically flat solutions, one imposes
on the metric functions
\begin{equation}
f|_{r=\infty}= h|_{r=\infty}= l|_{r=\infty}=1 \ , \ \ \
\omega|_{r=\infty}= 0
\ , \label{bc1a} \end{equation}
while for the dilaton one may impose
\begin{equation}
\Psi|_{r=\infty}=0
\ , \label{bc1b} \end{equation}
since any finite value of the dilaton field at infinity
can always be transformed to zero via
$\Psi \rightarrow \Psi - \Psi(\infty)$,
$r \rightarrow r e^{-\gamma \Psi(\infty)} $.

The gauge field in the asymptotic region can be obtained by a
gauge transformation of some gauge potential $A^{\infty}$ with 
gauge transformation matrix of the form Eq.~(\ref{gauge}).
It depends on the number of nodes, $k$, characterizing the respective gauge field 
functions and thus labeling the radially excitated solutions ($k>1$)
in analogy to the static spherically symmetric case.
In EYMD theory $A^{\infty}=0$ and $\Gamma = -2m\theta$, if $k$ is odd, resp. $\Gamma =0 $ if $k$ is even.
This yields the boundary conditions at $r=\infty$
\begin{equation}
B_1=B_2=H_1=H_3 =0 \ ,\ 1-H_2 =2m \ ,\ 1-H_4 =2\frac{\sin(m\theta)}{\sin\theta}\ , 
\ \ \ k \ {\rm odd} \ , 
\end{equation}
\begin{equation}
B_1=B_2=H_1=H_3 =0 \ ,\ 1-H_2 =0 \ ,\ 1-H_4 =0 \ , 
\ \ \ k \ {\rm even} \ . 
\end{equation}

\noindent {\sl  Boundary conditions at the horizon:} The event horizon of stationary 
black hole solutions resides at a surface of constant radial coordinate, $r=r_{\rm H}$,
and is characterized by the condition $f(r_{\rm H},\theta)=0$ \cite{Kleihaus:2000kg}.
Regularity at the horizon requires 
for the metric functions
\begin{equation}
f|_{r=r_{\rm H}}=
h|_{r=r_{\rm H}}=
l|_{r=r_{\rm H}}=0
\ , \ \ \ \omega|_{r=r_{\rm H}}=\omega_{\rm H}= {\rm const.}
\ , \label{bh2a} \end{equation}
and for the dilaton function 
\begin{equation}
\partial_r \Psi|_{r=r_{\rm H}} =0
\ , \label{bh2b} \end{equation}
while the magnetic gauge field functions satisfy
\begin{equation}
              H_1 |_{r=r_{\rm H}}= 0 \ , \ \ \
\partial_r H_2 |_{r=r_{\rm H}}= 
\partial_r H_3 |_{r=r_{\rm H}}= 
\partial_r H_4 |_{r=r_{\rm H}}= 0 \  
\ , \label{bh2d} \end{equation}
with the gauge condition $\partial_\theta H_1=0$ taken into account
\cite{Kleihaus:2000kg}.
The boundary conditions for the electric gauge field functions
follow from the requirement that
the electrostatic potential 
is constant at the horizon \cite{Kleihaus:2000kg,Kleihaus:2002ee,Kleihaus:2003sh}
\begin{equation}
\Phi_{\rm el} \frac{\tau_z}{2}= - \chi^\mu A_\mu |_{r=r_{\rm H}}
\  \label{esp0} \end{equation}
(where its dimensionless version is subsequently introduced
${\bar \Phi_{\rm el}} = \sqrt{4 \pi G}  \Phi_{\rm el}$
and the bar omitted). 
In terms of the horizon angular velocity $\Omega$,
one then finds the boundary conditions 
\begin{equation}
 B_1 |_{r=r_{\rm H}}  =   n \Omega \cos m \theta  \ , \ \ \
 B_2 |_{r=r_{\rm H}}  =  -n \Omega \sin m \theta  \ .
\label{bh2e}
\end{equation}

\noindent {\sl Boundary conditions along the symmetry axis:}
For the positive $z$-axis the boundary conditions are given by
\begin{eqnarray}
& &\partial_\theta f|_{\theta=0} =
   \partial_\theta h|_{\theta=0} =
   \partial_\theta l|_{\theta=0} =
   \partial_\theta \omega|_{\theta=0} = 0 \ ,
\label{bc4a} \end{eqnarray}
\begin{eqnarray}
& &\partial_\theta \Psi|_{\theta=0} = 0 \ ,
\label{bc4b} \end{eqnarray}
\begin{eqnarray}
 H_1|_{\theta=0}=H_3|_{\theta=0}=0 \ , \ \ \
   \partial_\theta H_2|_{\theta=0} =
   \partial_\theta H_4|_{\theta=0}  = 0 \ ,
\label{bc4c} \end{eqnarray}
\begin{eqnarray}
B_2|_{\theta=0}=0 \ , \ \ \ \partial_\theta  B_1|_{\theta=0}=0 \ ,
\label{bc4d} \end{eqnarray}
The analogous conditions hold on the negative $z$-axis.
We note, that for black hole solutions
reflection symmetry w.r.t.~the $xy$-plane is broken
via the boundary conditions of the time component of the gauge field.
In addition, regularity on the $z$-axis requires condition Eq.~(\ref{lm})
for the metric functions to be satisfied,
and regularity of the energy-momentum tensor on the $z$-axis requires
\begin{equation}
H_2|_{\theta=0}=H_4|_{\theta=0}
\ , \ \ \  H_2|_{\theta=\pi}=H_4|_{\theta=\pi} \ .
\label{h2h4} \end{equation}

\subsection{Global charges and horizon properties}

\noindent {\sl Global  charges:}
The global charges of the black hole solutions can be obtained from their
asymptotic expansions.
The asymptotic expansion for the metric function $f$ yields the 
dimensionless mass $M$
$$
f = 1-\frac{2 M}{r} +  O \left(\frac{1}{r^2}\right)\ ,
$$
the asymptotic expansions for the metric function $\omega$
yields the dimensionless angular momentum $J$
$$
\omega = \frac{2 J}{r^2} + O\left(\frac{1}{r^3}\right)\ ,
$$
and the asymptotic expansion for the dilaton function $\Psi$
yields the dimensionless dilaton charge
$$
\Psi = -\frac{D}{r} 
+ O\left(\frac{1}{r^2}\right)
\ .
$$
The asymptotic expansion of the gauge field yields the global
non-Abelian electromagnetic charges, $Q$ and $P$,
where the electric gauge field functions $B_1$ and $B_2$ contain
the non-Abelian electric charge $Q$ in their leading term,
$$
{B}_1 = \frac{Q \cos{\theta}}{r}
+ O\left(\frac{1}{r^2}\right)\ , \ \ \ \
{B}_2 = -(-1)^k \frac{Q \sin{\theta}}{r}
 + O\left(\frac{1}{r^2}\right)\ ,
$$
while the boundary conditions of the magnetic gauge field functions
guarantee that the dimensionless non-Abelian magnetic charge $P$ vanishes.
Although the non-Abelian global charges $Q$ and $P$ appear to be gauge dependent,
the modulus of the non-Abelian electric and magnetic charge, $|Q|$ and $|P|$,
can be given a gauge invariant definition \cite{Corichi:1999nw,Corichi:2000dm}
\begin{equation}
  |Q|= \frac{e}{4\pi}
 \oint \sqrt{\sum_i{\left({^*}F^i_{\theta\varphi}\right)^2}}
 d\theta d\varphi 
\ , \label{Qdelta} \end{equation}
\begin{equation}
 |P|= \frac{e}{4\pi}
 \oint \sqrt{\sum_i{\left(F^i_{\theta\varphi}\right)^2}} d\theta d\varphi
\ , \label{Pdelta} \end{equation}
where ${^*}F$ represents the dual field strength tensor,
and the integral is evaluated at spatial infinity.
In analogy to the electromagnetic case also a
non-Abelian magnetic moment can be introduced \cite{Kleihaus:2003sh}.

We note that the lowest order terms in the expansions, needed for
the global charges $M$, $J$, $D$, $Q$ and $P$,
do not involve non-integer powers.
However, the gauge field functions for odd $n$ possess
non-integer powers in the higher order terms,
while for even $n$ logarithmic terms
like $\log(r)/r^2$ arise already in the static limit \cite{Kleihaus:2002tc,Kleihaus:2003sh,Kleihaus:1999ia}.

\noindent {\sl Horizon properties:}
Expanding the solutions at the horizon in powers of
$\delta=\frac{r}{r_{\rm H}}-1$, the lowest order term of the metric functions
$f$, $h$ and $l$ is quadratic in $\delta$. Naming the respective coefficients
$f_2$, $H_2$ and $l_2$, the horizon properties can be expressed 
in terms of these expansion coefficients. The dimensionless area $A$ is given by
\begin{equation}
A = 2 \pi \int_0^\pi  d\theta \sin \theta
\frac{\sqrt{l_2 h_2}}{f_2} r_{\rm H}^2
\  \label{area} \end{equation}
and defines the area parameter $r_\Delta$ via \cite{Corichi:1999nw,Corichi:2000dm} 
\begin{equation}
A = 4 \pi r_\Delta^2 \ ,
\label{xDelta} \end{equation}
while the dimensionless entropy $S$ of the black hole is 
\begin{equation}
S = \frac{A}{4} \ .
\label{entro} \end{equation}
The surface gravity
\begin{equation}
\kappa_{\rm sg}^2 =
 - \frac{1}{2} (\nabla_\mu \chi_\nu)(\nabla^\mu \chi^\nu)
\ , \label{sgwald} \end{equation}
with Killing vector
$\chi$, Eq.~(\ref{chi}), reads
\begin{equation}
\kappa_{\rm sg} = \frac{f_2(\theta)}{r_{\rm H} \sqrt{h_2(\theta)}}
\ . \label{temp} \end{equation}
As required by the zeroth law of black hole mechanics,
$\kappa_{\rm sg}$ is constant on the horizon.
More formally this is seen by considering a null geodesic tangent 
to the Killing horizon together with the Einstein equations \cite{Wald:1984rg}.
The dimensionless temperature $T$ 
of the black holes is proportional to their surface gravity,
\begin{equation}
T = \frac{\kappa_{\rm sg}}{2 \pi} 
\ . \label{tempt} \end{equation}

The deformation of the horizon is obtained by comparing the
circumference of the horizon along the equator, 
$L_{\rm e}$,
and the circumference of the horizon along 
a great circle passing through the poles, 
$L_{\rm p}$,
\begin{equation}
L_{\rm e} = \int_0^{2 \pi} { d \vphi \left.
 \sqrt{ \frac{l}{f}} r \sin\theta
 \right|_{r=r_{\rm H}, \theta=\pi/2} } \ , \ \ \
L_{\rm p} = 2 \int_0^{ \pi} { d \theta \left.
 \sqrt{ \frac{h  }{f}} r
 \right|_{r=r_{\rm H}, \vphi={\rm const.}} }
\ , \label{lelp} \end{equation}
while further information 
resides in its Gaussian curvature $K$,
\begin{equation}
K(\theta) =  \frac{R_{\theta\vphi\theta\vphi}}{g_2} \ , \ 
g_2 =   g_{\theta\theta}g_{\vphi\vphi}-g_{\theta\vphi}^2
\ . \label{Gauss} \end{equation}
Its Euler characteristic $\chi_{\rm E} = 2$
shows that the horizon has the topology of a 2-sphere.

\noindent {\sl Mass formula:}
Electrovac black holes satisfy 
the Smarr formula \cite{Smarr:1972kt}
\begin{equation}
M = 2 TS + 2 \Omega J + \Phi_{\rm el} Q + \Phi_{\rm mag} P
\ , \label{smarr} \end{equation}
where $\Phi_{\rm mag}$ represents the horizon magnetic potential.
In the presence of a dilaton, an equivalent mass formula is given by
\begin{equation}
M = 2 TS + 2 \Omega J + 2 \Phi_{\rm el} Q + \frac{D}{\gamma} 
\ . \label{nasmarr} \end{equation}
Interestingly, in this form the mass formula remains valid
for non-Abelian fields \cite{Kleihaus:2002tc},
and represents a generalization of the non-Abelian mass formula
$M = 2 TS + D/\gamma$, obtained previously 
for static axially symmetric EYMD solutions \cite{Kleihaus:1997ic}.
Rotating EYMD black hole solutions generically 
carry non-Abelian electric charge, 
but they do not carry non-Abelian magnetic charge,
while they possess non-trivial non-Abelian magnetic fields
outside their horizon.
This magnetic field contribution to the mass is contained in 
the mass formula (\ref{nasmarr})
in the dilaton term $ {D}/{\gamma}$,
while it would be missing in  (\ref{smarr}).
The derivation of the mass formula  (\ref{nasmarr}) is based on the
Komar integrals for the global charges and the equations of motion
\cite{Kleihaus:2002tc,Kleihaus:2003sh}.

\subsection{Nonperturbative black holes}

The first results for rotating EYM black holes were obtained by applying perturbation theory
\cite{Volkov:1997qb,Brodbeck:1997ek}.
These perturbative considerations led to the prediction, that rotating EYM black holes should
come in three types: the first type carries angular momentum $J \ne 0$ and
electric charge $Q \ne 0$, the second type carries angular momentum $J \ne 0$ 
but has vanishing electric charge $Q = 0$, while the third  type has
vanishing angular momentum $J = 0$ but carries electric charge $Q \ne 0$
\cite{Volkov:1997qb,Brodbeck:1997ek}.

In the following we discuss the nonperturbative results obtained numerically
and compare with the perturbative predictions, addressing first the EYM 
and subsequently the EYMD black holes.

\subsubsection{EYM black holes}

Rotating EYM black hole solutions emerge from their static counterparts,
when a small value of the horizon angular velocity is imposed.
We recall that static EYM black holes
carry non-trivial magnetic gauge fields, but do not carry non-Abelian magnetic charge
(except for embedded Abelian black holes)
\cite{Volkov:1990sva,Bizon:1990sr,Kuenzle:1990is,Kleihaus:1997ic,Kleihaus:1997ws,Ibadov:2005rb}.
Moreover, their electric fields vanish identically (in SU(2) EYM theory).
These observations led to the non-Abelian baldness theorem, stating that all
static EYM black holes with finite non-Abelian charges are embedded Abelian
black holes
\cite{Galtsov:1989ip,Bizon:1992pi}.

The static black holes can be characterized by their mass and by the integers
$n$, $m$, and $k$. For given $n$ and $m$, 
the solutions form sequences
labelled by the node number $k$. 
The simplest sequence has $n=m=1$, corresponding to spherically symmetric
black holes \cite{Volkov:1990sva,Bizon:1990sr,Kuenzle:1990is}.
For $n>1$ or $m>1$ static black holes are found,
which possess only axial symmetry
\cite{Kleihaus:1997ic,Kleihaus:1997ws,Ibadov:2005rb},
showing that Israel's theorem does not generalize to non-Abelian theories.
The EYM black holes can have arbitrary size.
But only for sufficiently small black holes
the non-Abelian fields contribute significantly to the mass.
Since the mass is their only global charge,
the static EYM black holes violate uniqueness.

In contrast to the static EYM black holes,
their rotating generalizations do carry non-Abelian electric gauge fields.
While this is expected, and well-known from the Abelian case,
where a magnetically charged RN black hole turns into 
a KN black hole with an induced electric dipole moment,
when set into rotation,
it is surprizing, that the rotation of the neutral EYM black holes
with non-trivial non-Abelian magnetic fields
induces a non-Abelian electric charge
\cite{Volkov:1997qb,Brodbeck:1997ek}.

Notably, however, all known rotating EYM black holes possess
a non-Abelian electric charge 
\cite{Kleihaus:2000kg,Kleihaus:2002tc}.
Thus they all represent black holes of the first type, $J \ne 0$ and $Q \ne 0$,
predicted by perturbation theory for rotating EYM black holes
\cite{Volkov:1997qb,Brodbeck:1997ek}.
The second type with $J \ne 0$ and $Q = 0$, as well as the third  type with
$J = 0$ and $Q \ne 0$ do not appear to be realized non-perturbatively.
Since the non-Abelian magnetic charge
of the rotating EYM black holes is identically zero,
they carry three gobal charges: $M$, $J$ and $Q$.
Like their static counterparts, for given integers $n$ and $m$
they form sequences labelled by the node number $k$. 

The set of equations to be solved consists of 
ten coupled non-linear elliptic partial differential equations.
It is convenient to employ compactified dimensionless coordinates,
mapping spatial infinity to the finite value $\tilde r=1$,
where
\begin{equation}
\tilde r = 1-\frac{r_{\rm H}}{r}
\ . \label{barx2} \end{equation}
In the numerical scheme \cite{Schoenauer},
based on the Newton-Raphson method,
the equations are discretized on a non-equidistant grid in $\tilde r$ and  $\theta$,
where typical grids used have sizes $100 \times 50$
(see e.g.~\cite{Kleihaus:2000kg,Kleihaus:2002tc} for further details
on the numerical procedure.)

Rotating black holes can be obtained by starting from a static black hole and
imposing a small but finite value of the horizon angular velocity $\Omega$
via the boundary condition for the function $\omega$,
i.e., $\omega_{\rm H}>0$.
In the simplest case one starts from a fundamental EYM
black hole solution which is static and spherically symmetric
($n=m=1$) and possesses a single node ($k=1$). 
The rotation then induces non-trivial  
functions for $\omega$, $H_1$, $H_3$, $B_1$, and $B_2$,
and the rotating black holes carry a non-Abelian electric gauge field
and an associated electric charge 
\cite{Volkov:1997qb,Kleihaus:2000kg,Kleihaus:2002tc}.

For a fixed isotropic horizon radius $r_{\rm H}$ the increase of 
the horizon angular velocity 
and thus of $\omega_{\rm H}$ yields a branch of 
rotating black holes,
which extends up to a maximal value $\omega^{\rm max}_{\rm H}$,
where a second branch is met, that
extends towards $\omega_{\rm H}=0$.
Along both branches the mass $M$, the angular momentum $J$, and the
non-Abelian electric charge $Q$ increase monotonically.
This is seen in Figs.~1a-c for three values of the isotropic horizon radius,
$r_{\rm H}=0.1$, 0.5 and 1.
(We recall, that the occurrence of the branches and the maximum
value $\omega^{\rm max}_{\rm H}$ is a feature that derives from the
use of isotropic coordinates.)

\begin{figure}[p!]
\mbox{
\includegraphics[height=.225\textheight, angle =0]{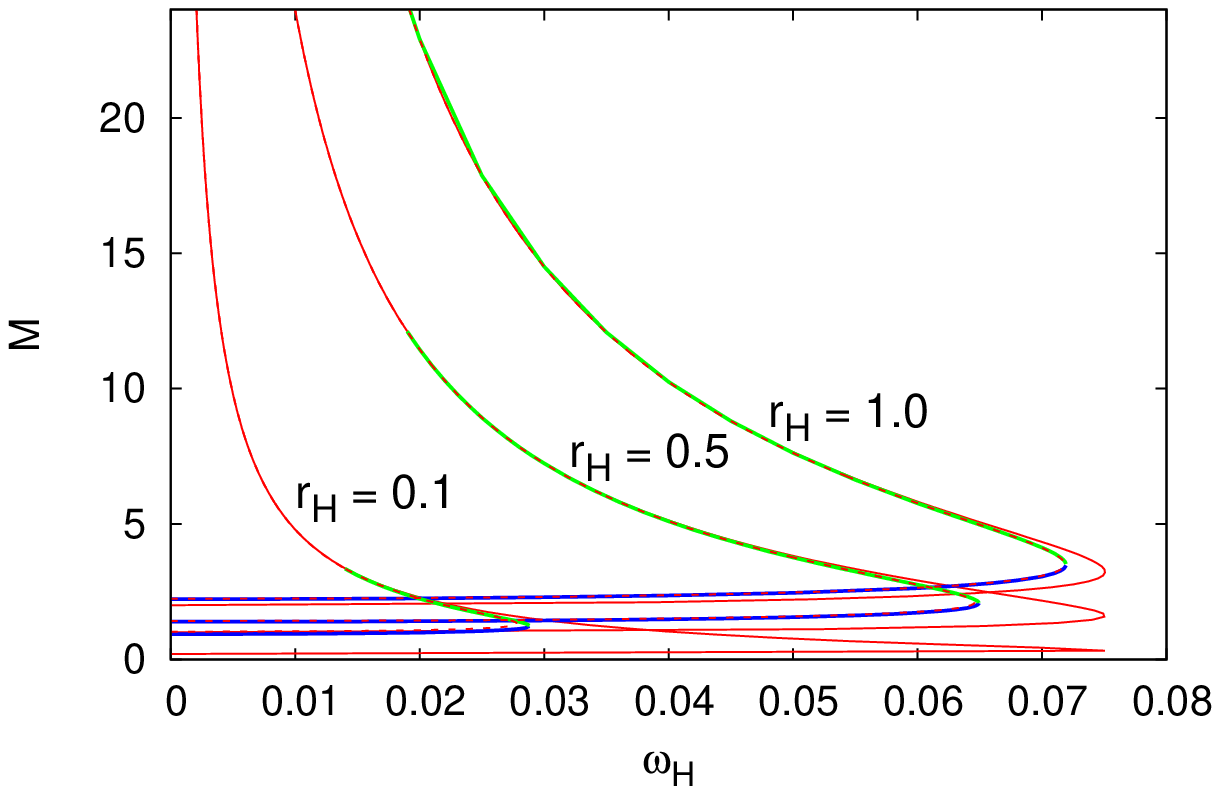}
\hspace*{0.5cm}
\includegraphics[height=.225\textheight, angle =0]{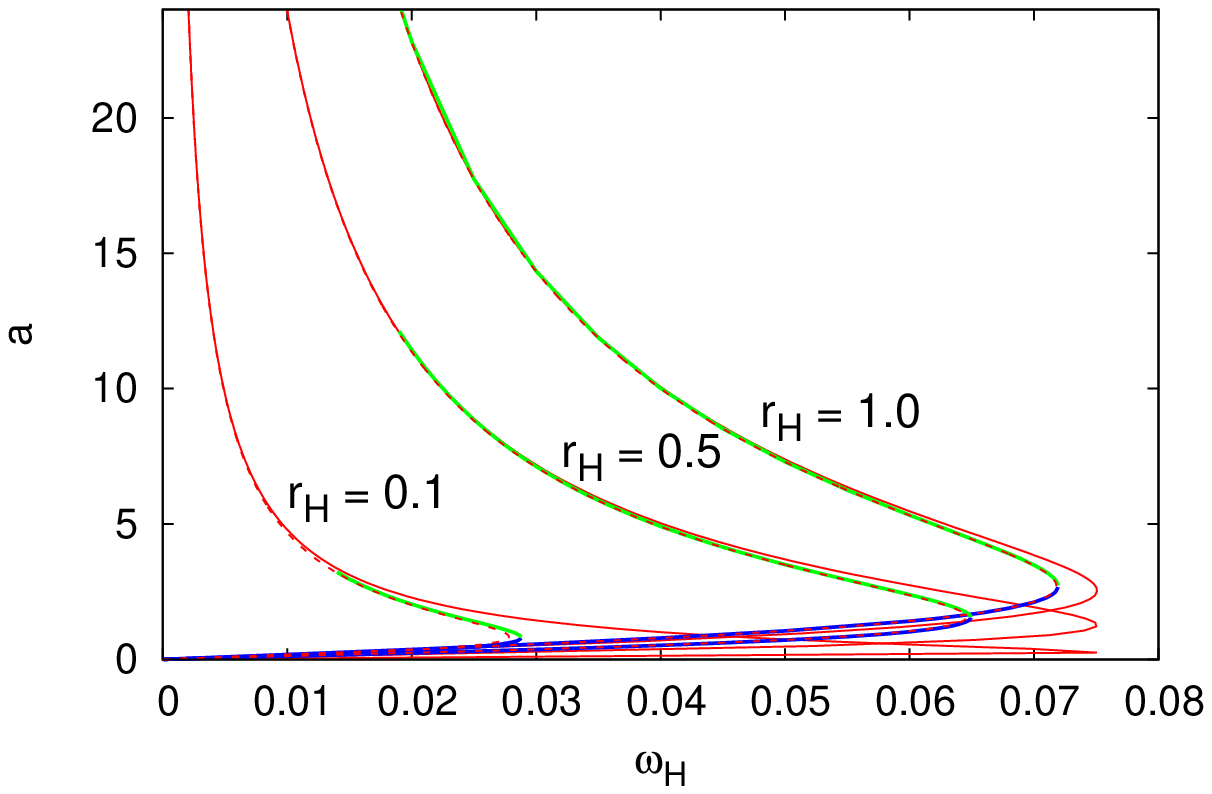}
}
\mbox{
\includegraphics[height=.225\textheight, angle =0]{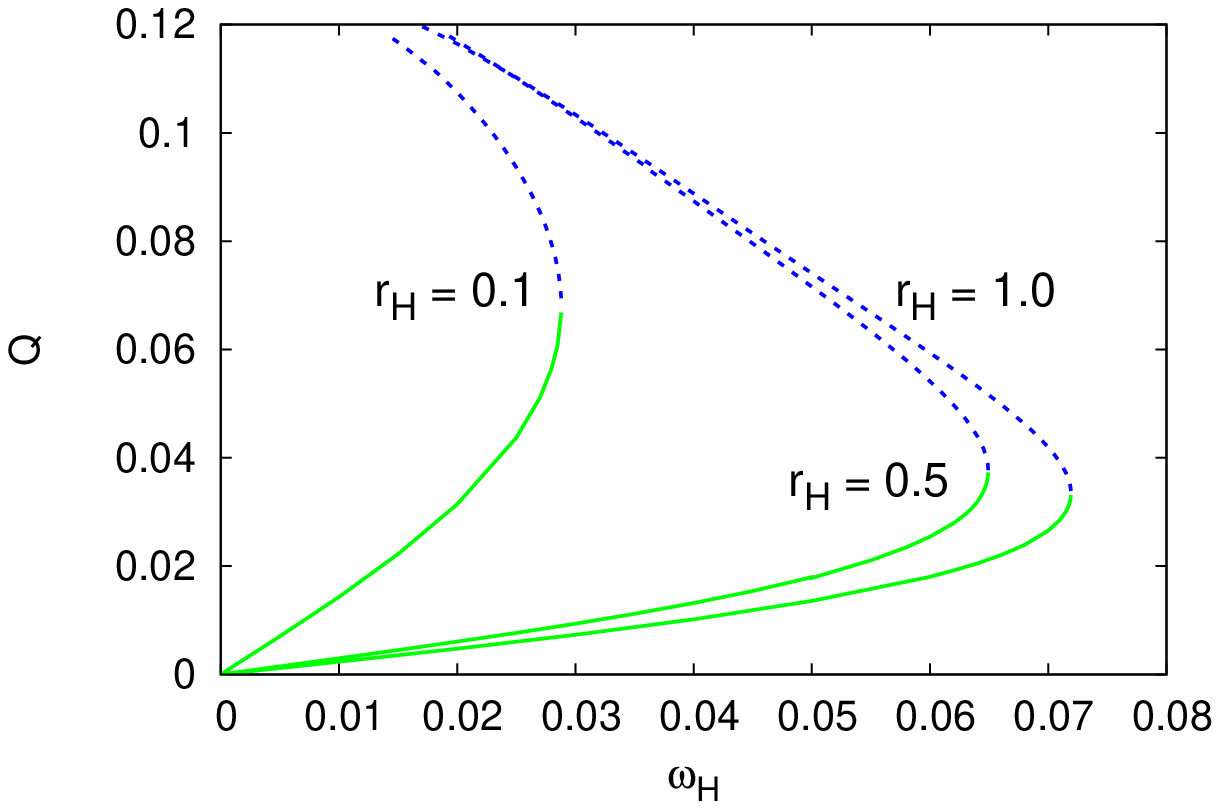}
\hspace*{0.5cm}
\includegraphics[height=.225\textheight, angle =0]{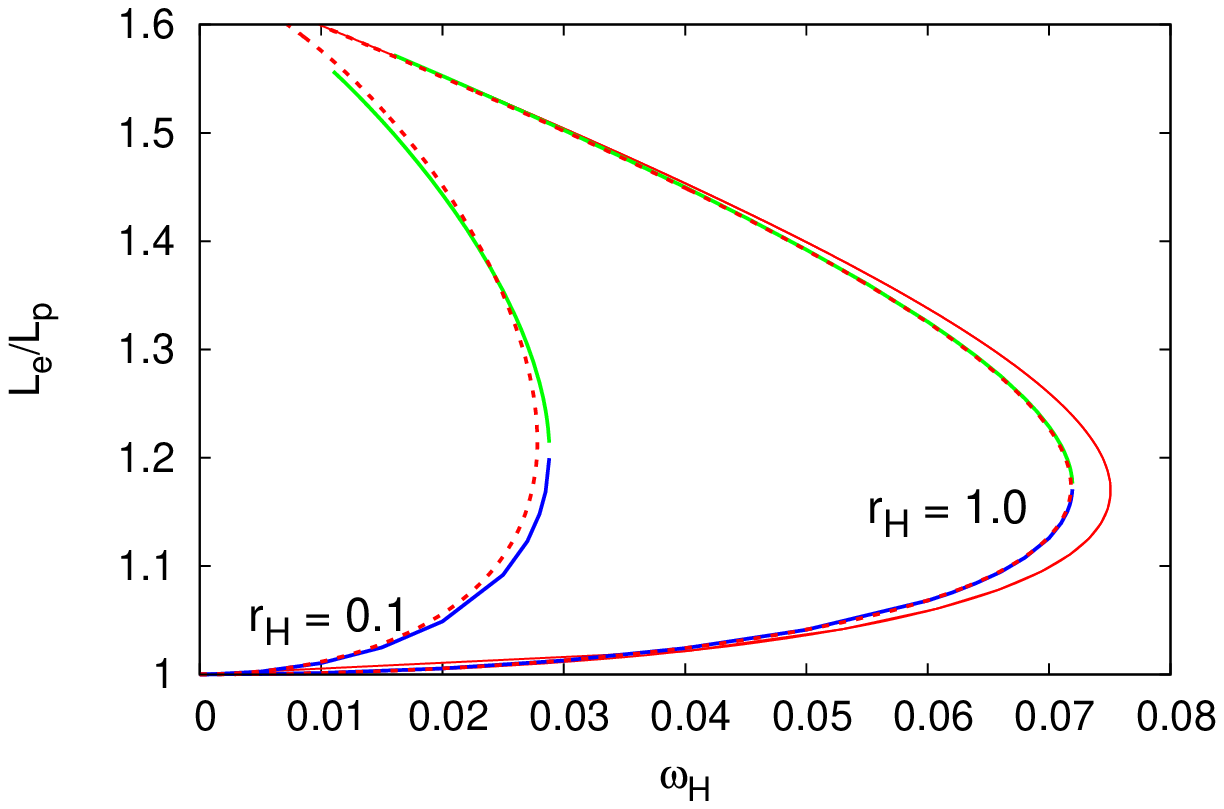}
}
\mbox{
\includegraphics[height=.225\textheight, angle =0]{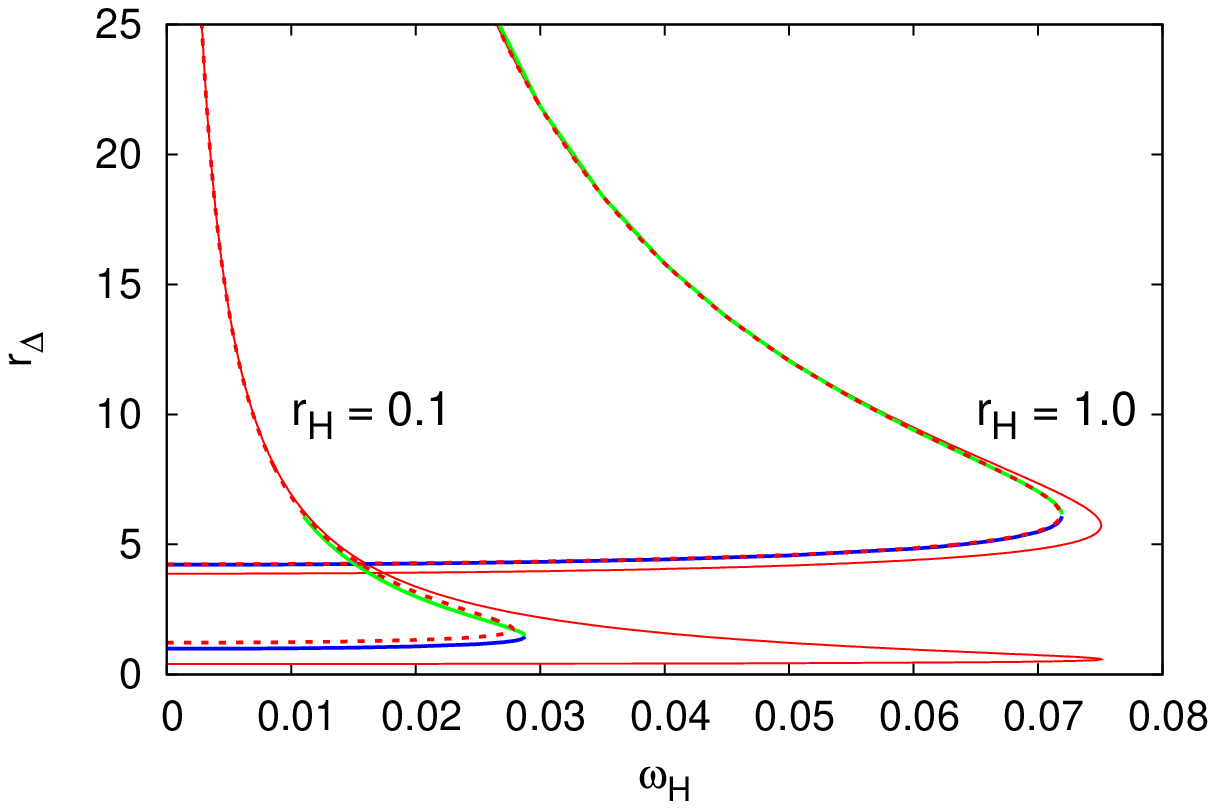}
\hspace*{0.5cm}
\includegraphics[height=.225\textheight, angle =0]{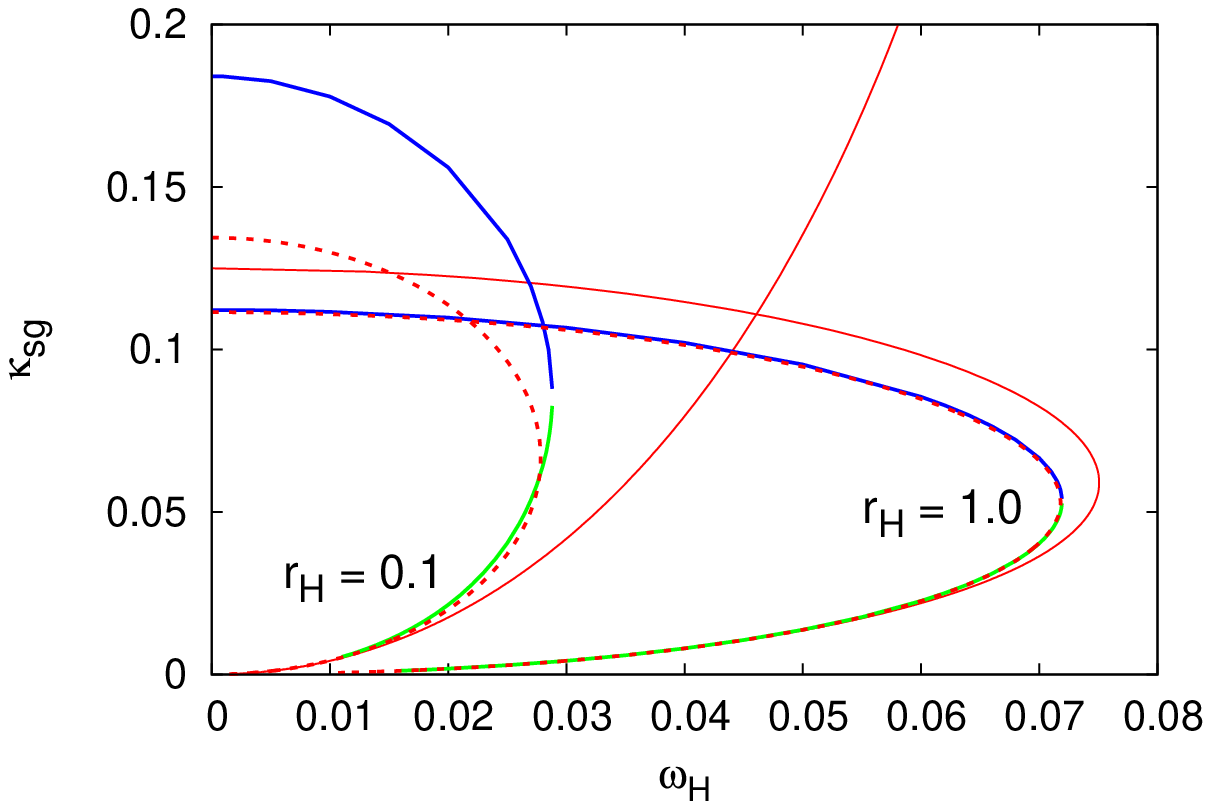}
}
\vspace{-0.5cm}
\caption{
EYM black holes with $n=1$, $m=1$, $k=1$ are compared to 
Kerr black holes (thin solid) and
Kerr-Newman black holes (dotted) with $Q=0$ and $P=1$:
a) The dimensionless mass $M$ is shown as a function of $\omega_{\rm H}$ for
$r_{\rm H} = 1.0 $, $0.5$ and $0.1$.
b) Same as a) for the specific angular momentum $a=J/M$ .
c) Same as a) for the electric charge  $Q$.
d) The ratio of the horizon circumferences $L_e/L_p$ is shown 
as a function of $\omega_{\rm H}$ for
$r_{\rm H}=1.0$ and $r_{\rm H}=0.1$ on the lower branch (solid)
and on the upper branch (dashed).
e) Same as d) for the area parameter $r_\Delta$.
f) Same as d) for the surface gravity $\kappa_{\rm sg}$.
}
\end{figure}


Whereas the mass $M$ and the angular momentum per unit mass $a$
increase without bound  along the upper branch, 
and diverge with $\omega_{\rm H}^{-1}$ in the
limit $\omega_{\rm H} \rightarrow 0$,
the non-Abelian electric charge $Q$ remains always small,
approaching a finite limiting value
$Q_{\rm lim} \approx 0.124$,
independent of the isotropic horizon radius $r_{\rm H}$.
This limit can be seen to arise when solving the gauge field equations
in the background of an extremal Kerr black hole
\cite{Kleihaus:2002tc}.

In Figs.~1a-b also
the mass and the angular momentum of two families of embedded Abelian solutions
are shown for comparison: i) the Kerr solutions and 
ii) the Kerr-Newman solutions 
with $Q^2+P^2=1$ and
the same horizon radii.
Interestingly, both the mass and the angular momentum of the EYM black holes
which carry a small electric charge $Q$ and no magnetic charge,
are rather close to the mass and the angular momentum of the embedded
Kerr-Newman solutions with $Q^2+P^2=1$.
Keeping in mind that for the rotating EYM black holes
the induced electric fields are small with $Q^2 \ll 1$, 
we conclude that their properties
can be mimicked rather well by 
magnetically charged Kerr-Newman black holes with $P=1$.
When considering the mass and angular momentum of EYM black holes
with higher node numbers $k>1$, one observes that
the corresponding quantities are still closer to those of the
Kerr-Newman solutions with unit magnetic charge. 

When a static spherically symmetric black hole
is set into rotation its horizon deforms.
The deformation of the horizon can be measured by the ratio of
the circumference of the horizon along the equator, $L_e$,
and the circumference of the horizon along the poles, $L_p$,
Eqs.~(\ref{lelp}).
The ratio $L_e/L_p$ is shown in Fig.~1d for EYM black holes
with $n=m=k=1$.
It grows monotonically along both branches.
On the upper branch 
in the limit $\omega_{\rm H} \to 0$, 
the ratio tends to the value  $L_e/L_p \approx 1.645$,
which corresponds to the value of an extremal Kerr black hole.
On the lower branch the ratio $L_e/L_p$ assumes the value one
in the limit $\omega_{\rm H} \to 0$,
as it must for a static spherically symmetric black hole.

The horizon size is exhibited in Fig.~1e, 
where the area parameter $r_\Delta$ is shown versus $\omega_{\rm H}$
for EYM black holes with fixed isotropic horizon radii
$r_{\rm H}=0.1$ and $r_{\rm H}=1$.
The horizon area grows monotonically along both branches,
and follows closely the Kerr-Newman area
for a  black hole with unit charge.
The surface gravity $\kappa_{\rm sg}$ of EYM black holes is shown in Fig.~1f.
Starting from the value of the corresponding 
static EYM black hole,
it decreases monotonically along both branches
and tends to zero in the limit $\omega_{\rm H} \rightarrow 0$
on the upper branch, the value assumed by extremal black holes.

\begin{figure}\centering
a\mbox{\epsfysize=7.5cm  \epsffile{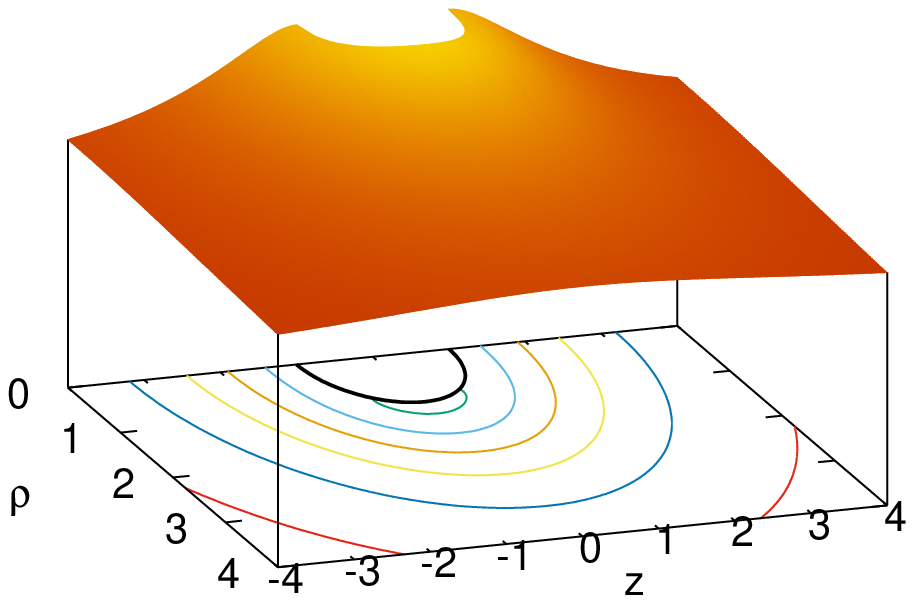}  }
\vspace{1.cm}\\
\begin{tabular}{ccc}
b & c & d \\
\mbox{\hspace*{-15mm}\epsfysize=5.cm\epsffile{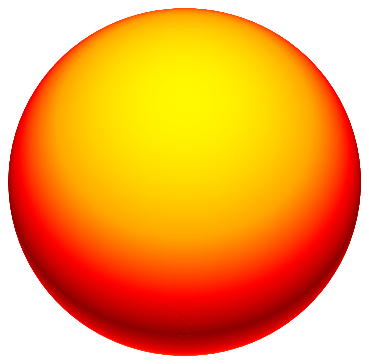}\hspace*{-15mm}}
&
\raisebox{0.mm}{\hspace*{-15mm}\epsfysize=5.0cm\epsffile{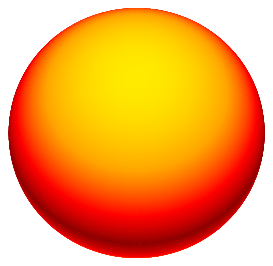}\hspace*{-15mm}}
&
\raisebox{0.0mm}{\hspace*{-15mm}\epsfysize=5.0cm\epsffile{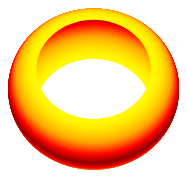}\hspace*{-15mm}}
\\
$\varepsilon = 0.0006$ &
$\varepsilon = 0.0009$ &
$\varepsilon = 0.0012$
\end{tabular}
\caption{
The component $\varepsilon=-T_0^0$ of the stress energy tensor is shown as a
function of the coordinates $\rho = r \sin\theta$, $z=r\cos\theta$
for $k=1$, $n=m=1$, $r_{\rm H}=1.0$, $\omega_{\rm H} = 0.04$ on the lower branch.
}
\end{figure}
    
\begin{figure}\centering
\mbox{\epsfysize=7.5cm  \epsffile{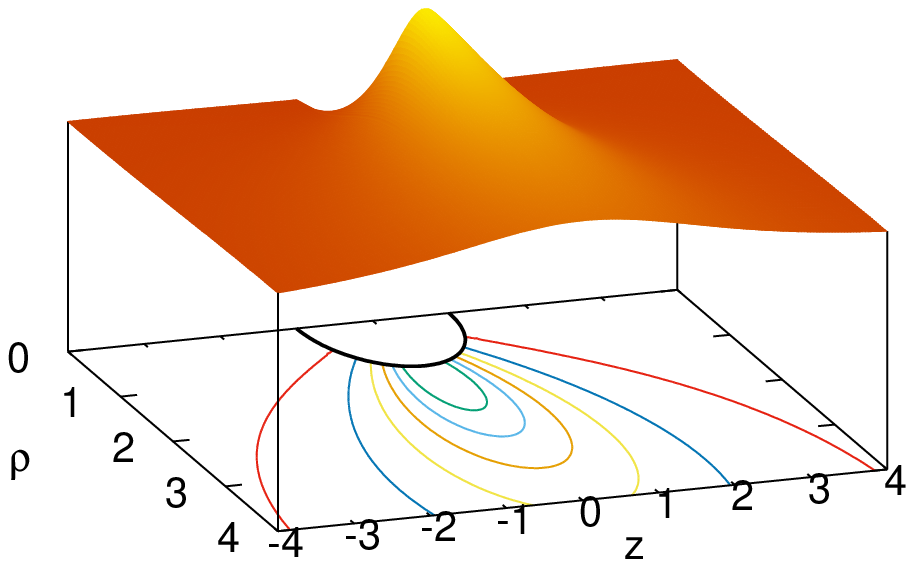}  }
\vspace{1.cm}\\
\begin{tabular}{ccc}
b & c & d \\
\mbox{\hspace*{-15mm}\epsfysize=5cm\epsffile{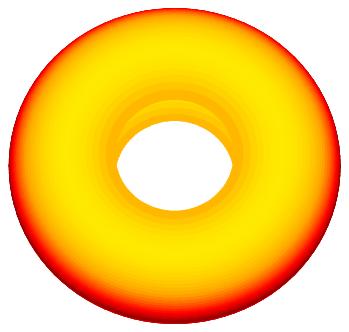}\hspace*{-15mm}}
&
\raisebox{0.mm}{\hspace*{-15mm}\epsfysize=5cm\epsffile{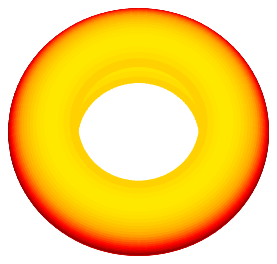}\hspace*{-15mm}}
&
\raisebox{0.mm}{\hspace*{-15mm}\epsfysize=5cm\epsffile{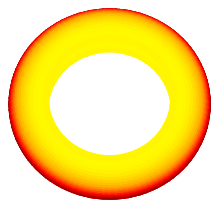}\hspace*{-15mm}}
\\
$\varepsilon = 0.00004$ &
$\varepsilon = 0.00005$ &
$\varepsilon = 0.00006$
\end{tabular}
\caption{
The same as Fig.~2 on the upper branch.
}
\end{figure}
      
To illustrate the distribution of the stress-energy of the non-Abelian fields,
we exhibit in Figs.~2 and 3 the component $\varepsilon = - T^0_0$
of the stress energy tensor. In the static case this corresponds to the
energy density of the black holes.
Due to the rotation, $\varepsilon$ is angle-dependent.
The maximum of $\varepsilon$ resides on the $\rho$-axis at the horizon,
as seen in the two representative examples.
Fig.~2a shows a 3-dimensional plot of $\varepsilon$
versus the coordinates $\rho = r \sin \theta$ and
$z= r \cos \theta$ together with a contour plot,
while Figs.~2b-d show surfaces of constant $\varepsilon$.
These surfaces are flattened at the poles, and bulge out in the equatorial plane.
For the largest values of $\varepsilon$ the horizon
can be seen in the pole region (see Fig.~2d).

Analogously,
Figs.~3a-d exhibit $\varepsilon$ for a
black hole solution with the same parameters
but on the upper branch.
Here $\varepsilon$ is much stronger deformed and shows a
large peak on the $\rho$-axis. This implies torus-shaped 
surfaces of constant $\varepsilon$,
with the horizon residing at the center the torus.
Only further away from the black hole horizon
the surfaces of constant $\varepsilon$ are again ellipsoidal.

The inclusion of rotation is not expected to stabilize these EYM black holes,
which in the static case have been shown to possess $2k$ unstable modes
(see e.g.~\cite{Volkov:1998cc} and references therein).
Also the generalized EYM black holes with $n>1$ or $m>1$ are expected
to be unstable, although a stability analysis here would be much more
involved. Their further properties will be addressed in the next
subsection, since the presence of a dilaton does not change most of their
basic features.

\subsubsection{EYMD black holes}

\begin{figure}[p!]
\mbox{
\includegraphics[height=.27\textheight, angle =0]{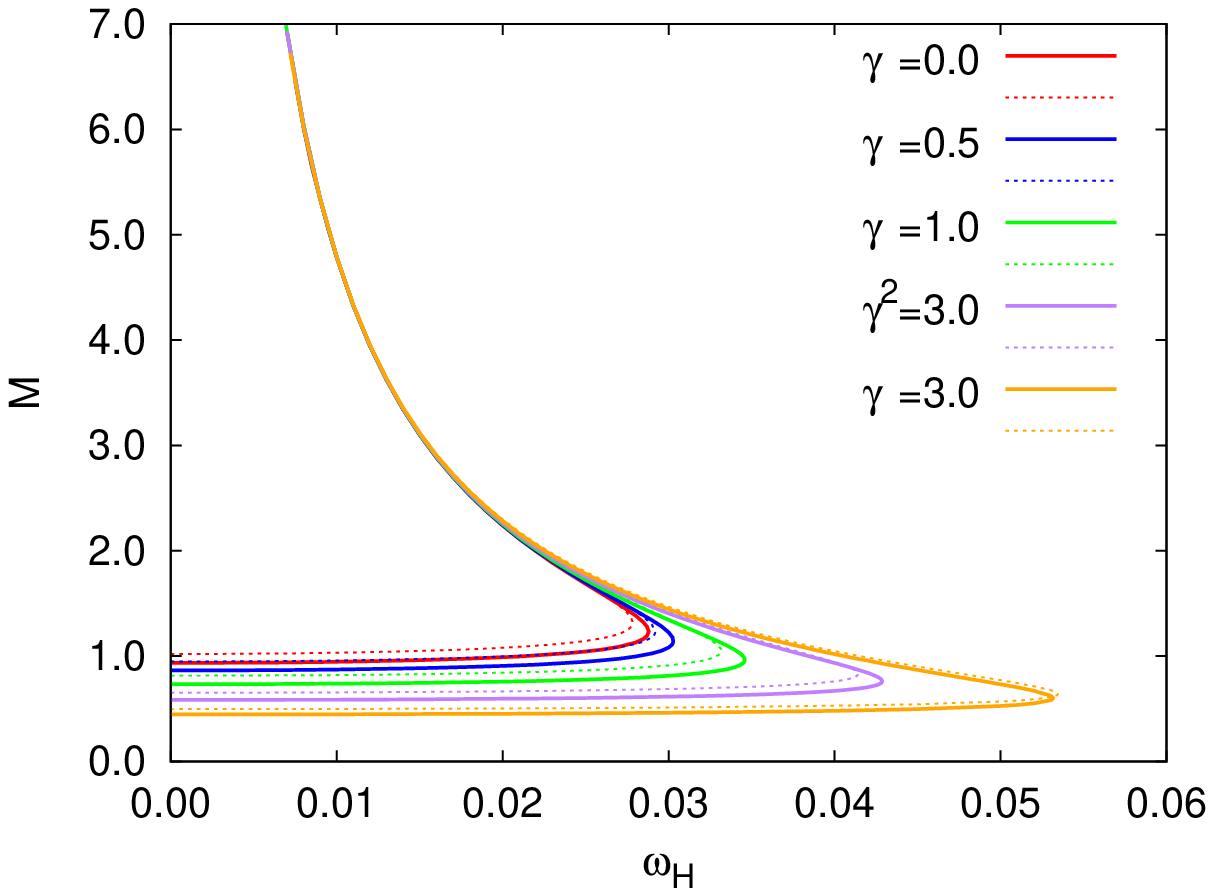}
\hspace*{0.5cm}
\includegraphics[height=.27\textheight, angle =0]{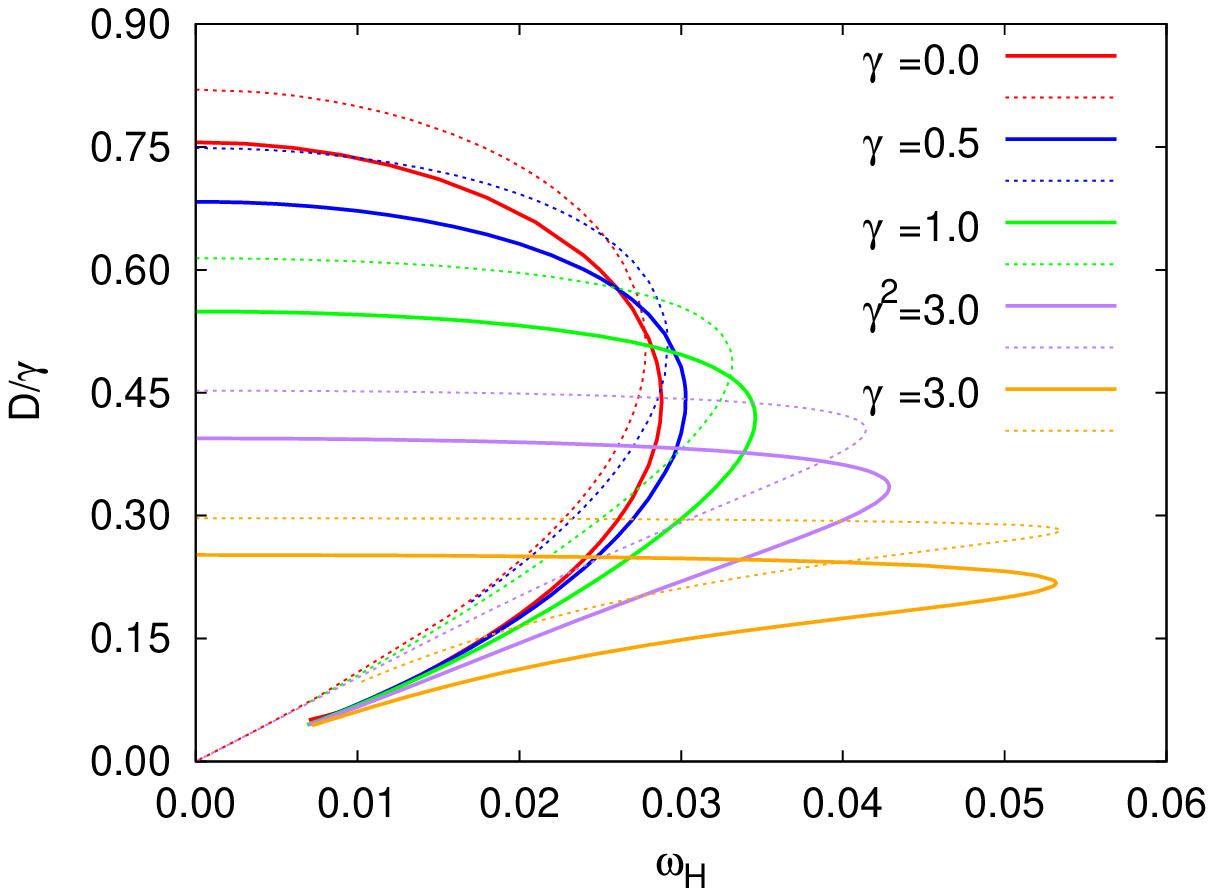}
}
\mbox{
\includegraphics[height=.27\textheight, angle =0]{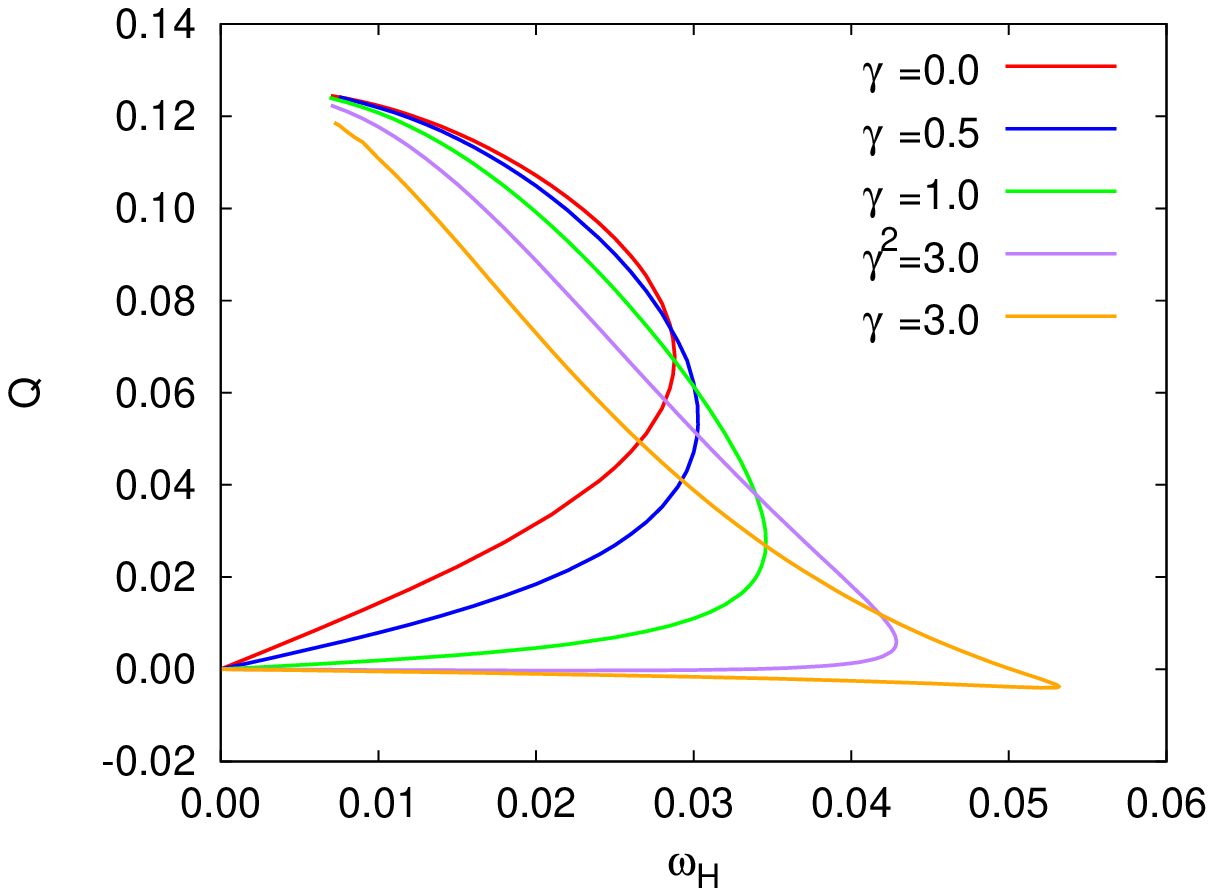}
\hspace*{0.5cm}
\includegraphics[height=.27\textheight, angle =0]{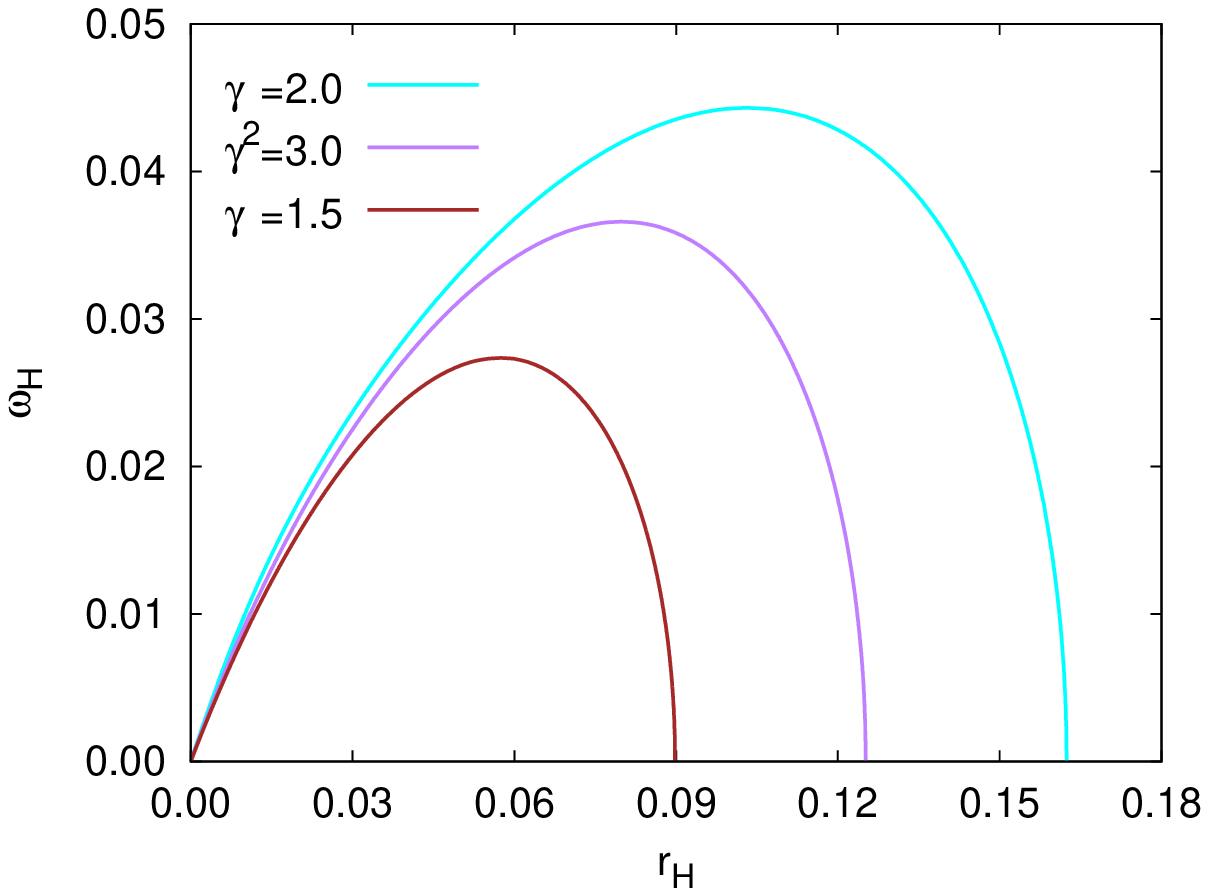}
}
\mbox{
\includegraphics[height=.27\textheight, angle =0]{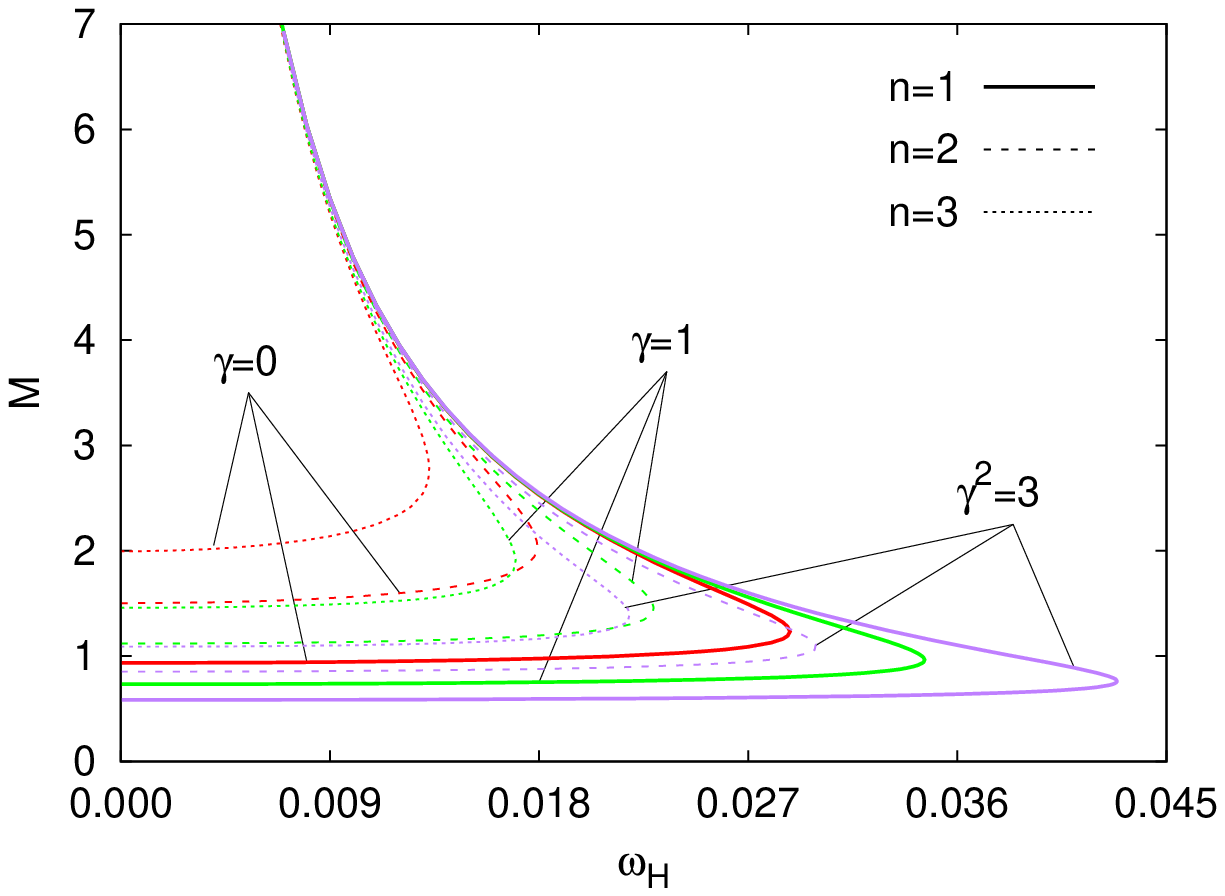}
\hspace*{0.5cm}
\includegraphics[height=.27\textheight, angle =0]{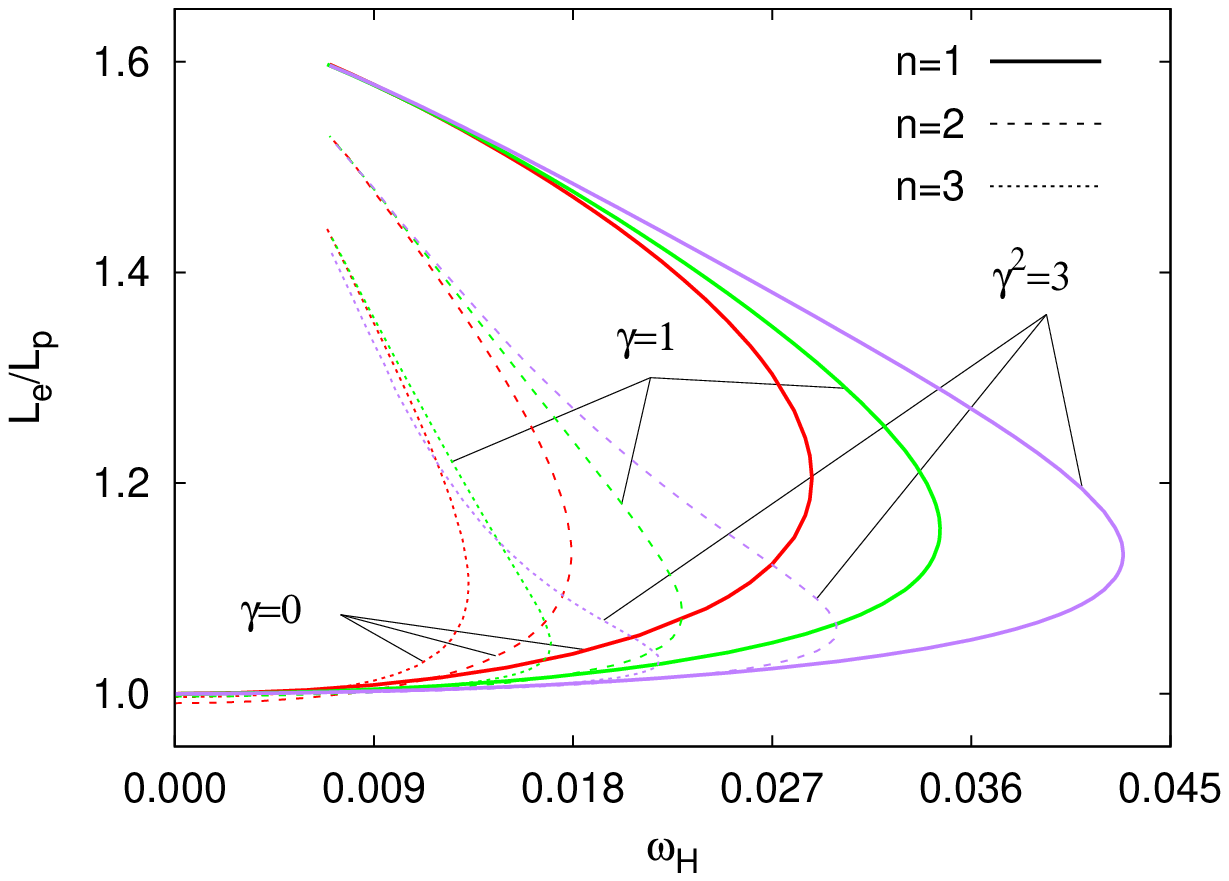}
}
\vspace{-0.5cm}
\caption{
Rotating EYMD black holes:
a) The dimensionless mass $M$ is shown as a function of $\omega_{\rm H}$ for
$n=1$, $m=1$, $k=1$,
$r_{\rm H} =  0.1$, and several values of $\gamma$.
Also shown are EMD
black holes (dotted) with $Q=0$ and $P=1$.
b) Same as a) for the relative dilaton charge $D/\gamma$.
c) Same as a) for the non-Abelian electric charge $Q$.
d) Curves of vanishing non-Abelian electric charge for $n=1$, $m=1$, $k=1$,
and several values of $\gamma$ in the $r_{\rm H}$-$\omega_{\rm H}$ plane.
e) The dimensionless mass $M$ is shown as a function of $\omega_{\rm H}$ for
$n=1,2,3$, $m=1$, $k=1$,
$r_{\rm H} =  0.1$, and several values of $\gamma$.
f) Same as e) for the ratio of the horizon circumferences $L_e/L_p$.
}
\label{fig4}
\end{figure}

Rotating EYMD black holes have many features in common with their
EYM counterparts, from which they arise when the dilaton  coupling constant $\gamma$
is increased from zero \cite{Kleihaus:2002tc,Kleihaus:2003sh}.
In the following we address the influence of the dilaton
on the properties of these black hole solutions.
A nice surprize here is the occurrence of black hole solutions of the second type
predicted by perturbation theory \cite{Volkov:1997qb,Brodbeck:1997ek},
i.e., black holes with angular momentum $J \ne 0$ but vanishing electric charge $Q = 0$,
in addition to the black holes of the first type,
which carry angular momentum $J \ne 0$ and
electric charge $Q \ne 0$, and which represent the only type encountered so far
in EYM theory.

Let us start the discussion again with the global charges and
consider the rotating generalizations of the fundamental 
static spherically symmetric EYMD black holes with
$n=m=k=1$ first. Figs.~4a-c illustrate some of their properties.
The dependence of the mass on the dilaton coupling constant $\gamma$
is demonstrated in Fig.~4a, where the two branches of black holes
that arise for fixed isotropic horizon radius and varying horizon
angular velocity resp.~$\omega_{\rm H}$ are exhibited for
several values of $\gamma$. 
The mass of the EYMD black holes shows the expected behavior:
it follows rather closely the mass of the Abelian Einstein-Maxwell-Dilaton (EMD)
black holes with charge $P=1$. Accordingly,
the maximum value $\omega^{\rm max}_{\rm H}(x_{\rm H},\gamma)$
of these EYMD black holes
increases with increasing $\gamma$.
We recall, that rotating EMD black holes for arbitrary values of $\gamma$ must 
also be obtained numerically \cite{Kleihaus:2003df}.
Only for the Kaluza-Klein value of $\gamma = \sqrt{3}$ they are known analytically
\cite{Frolov:1987rj,Rasheed:1995zv} (and of course for $\gamma=0$).

Since the specific angular momentum $a=J/M$ has a very similar behavior
to the mass, also following rather closely the respective EMD curves,
we do not exhibit it here.
Instead we now focus on the new features of the EYMD black holes.
First of all, they carry a new global charge, the dilaton charge $D$.
The relative dilaton charge $D/\gamma$ is exhibited in Fig.~4b
for the same set of black hole solutions.
Both for the non-Abelian and Abelian black holes,
$D/\gamma$ tends to zero on the upper branches,
since the dilaton is sourced by the matter fields, and these
become rather irrelevant for these large black holes.
Note, that
the limiting $D/\gamma=0$ curves in the figure are extracted from the Smarr formula
(\ref{nasmarr}), which for finite values of $\gamma$ represents a good check of the
quality of the numerical calculations.

The non-Abelian electric charge $Q$ is exhibited in Fig.~4c.
While $Q$ increases monotonically 
along both branches, when $\gamma < 1.15$
(including the EYM case $\gamma=0$),
for $\gamma > 1.15$ (and $r_{\rm H}$ sufficiently small)
it first decreases along the lower branch
until some minimal value is reached.
Thus in a certain parameter range 
the induced charge 
of the EYMD black holes is negative, while its magnitude remains always small.
For EYM black holes $Q$ approaches the finite (positive) limiting value
$Q_{\rm lim} \approx 0.124$ on the upper branch for
$\omega_{\rm H} \to 0$,
independent of the isotropic horizon radius $r_{\rm H}$ \cite{Kleihaus:2002ee},
which seems to hold as well for EYMD black holes \cite{Kleihaus:2002tc,Kleihaus:2003sh}.

It is most remarkable that, depending on the magnitude of the horizon angular velocity,
the non-Abelian electric charge $Q$
can change sign in the presence of the dilaton field \cite{Kleihaus:2002tc,Kleihaus:2003sh}.
In contrast,
for EYM black holes the sign of the non-Abelian electric charge $Q$
depends only on the direction of the rotation
\cite{Kleihaus:2002ee}.
Thus rotating EYMD black holes can have vanishing non-Abelian electric charge $Q$
for combinations of the dilaton coupling constant $\gamma$,
the horizon radius $r_{\rm H}$ and the horizon angular velocity 
$\Omega=\omega_{\rm H}/r_{\rm H}$.

A set of these special rotating $Q=0$ EYMD black holes
is exhibited in Fig.~4d for $\gamma=1.5$, $\sqrt{3}$, and $2$.
Rotating EYMD black holes with negative $Q$ would be found below
those curves.
We note that
in the limit $\gamma \rightarrow \gamma_{\rm min} \approx 1.15$ the 
curves degenerate to a point, i.e., rotating $Q=0$ solutions 
exist only above $\gamma_{\rm min}$.
An interesting consequence of the vanishing of the non-Abelian electric charge
is the disappearance of the non-integer exponents in the
asymptotic expansions of these $Q=0$ EYMD black holes:
they exhibit a simple power-law behavior.

The horizon properties of the rotating EYMD black holes are similar to those
of their EMD counterparts (with $P=1$).
Analogous to Kerr-Newman black holes,
the Gaussian curvature at the horizon can become negative 
for rapidly rotating EYMD and EMD black holes.
For specific angular momentum $a>0$, 
the Gaussian curvature increases monotonically from the pole 
to the equator. When $a$ is sufficiently large,
the Gaussian curvature becomes negative in the vicinity
of the pole. An isometric embedding can then no longer be
performed completely in Euclidean space,
but the region with negative curvature
must be embedded in pseudo-Euclidean space 
\cite{Smarr:1972kt,Kleihaus:2003sh}.


Fig.~4 also addresses rotating EYMD black holes with higher
winding numbers, $n>1$.
These black holes retain axial symmetry in the static limit
\cite{Kleihaus:1997ic,Kleihaus:1997ws}, whereas the
$n=1$ black holes become spherically symmetric
when the rotation is turned off.
Still, many features of the $n>1$ EYMD black holes,
and analogously of the EYM black holes,
agree with those of their $n=1$ counterparts.
This is, for instance, seen in Fig.~4e where we exhibit the mass
for black holes with horizon radius $r_{\rm H}=0.1$, 
winding numbers $n=1-3$, (retaining $m=1$), and node number $k=1$
for the dilaton coupling constants $\gamma=0$, $1$, and $\sqrt{3}$
versus $\omega_{\rm H}$.

While the relative dilaton charge $D/\gamma$ always
tends to zero on the upper branch when $\omega_{\rm H} \rightarrow 0$, 
it approaches zero from below for winding number $n=3$,
and thus negative values of $D$ arise.
As in the case $n=1$, also for the higher $n$
the non-Abelian electric charge $Q$ tends to a finite limiting value,
which increases with increasing $n$, leading to significantly higher
values of $Q$.
A comparison of the global charges of these non-Abelian black holes
with those of the corresponding Abelian black holes with charge $P=n$
shows, that whereas
the discrepancies between Abelian and non-Abelian solutions 
become larger with increasing winding number $n$,
the Abelian ones still exhibit the same basic pattern.

Also the horizon properties of the $n>1$ black holes follow
the pattern of their Abelian counterparts.
To illustrate the $n$-dependence, we exhibit in Fig.~4f
the deformation of the horizon in terms of the ratio
of equatorial to polar circumferences, $L_{\rm e}/L_{\rm p}$
for the same set of EYMD black holes as in Fig.~4e.
Along the upper branch the ratio $L_{\rm e}/L_{\rm p}$
becomes (rather) independent of $\gamma$,
while retaining a distinct $n$-dependence
in the approach towards the common limiting Kerr value.

\begin{figure}[t]
\mbox{
\includegraphics[height=.245\textheight, angle =0]{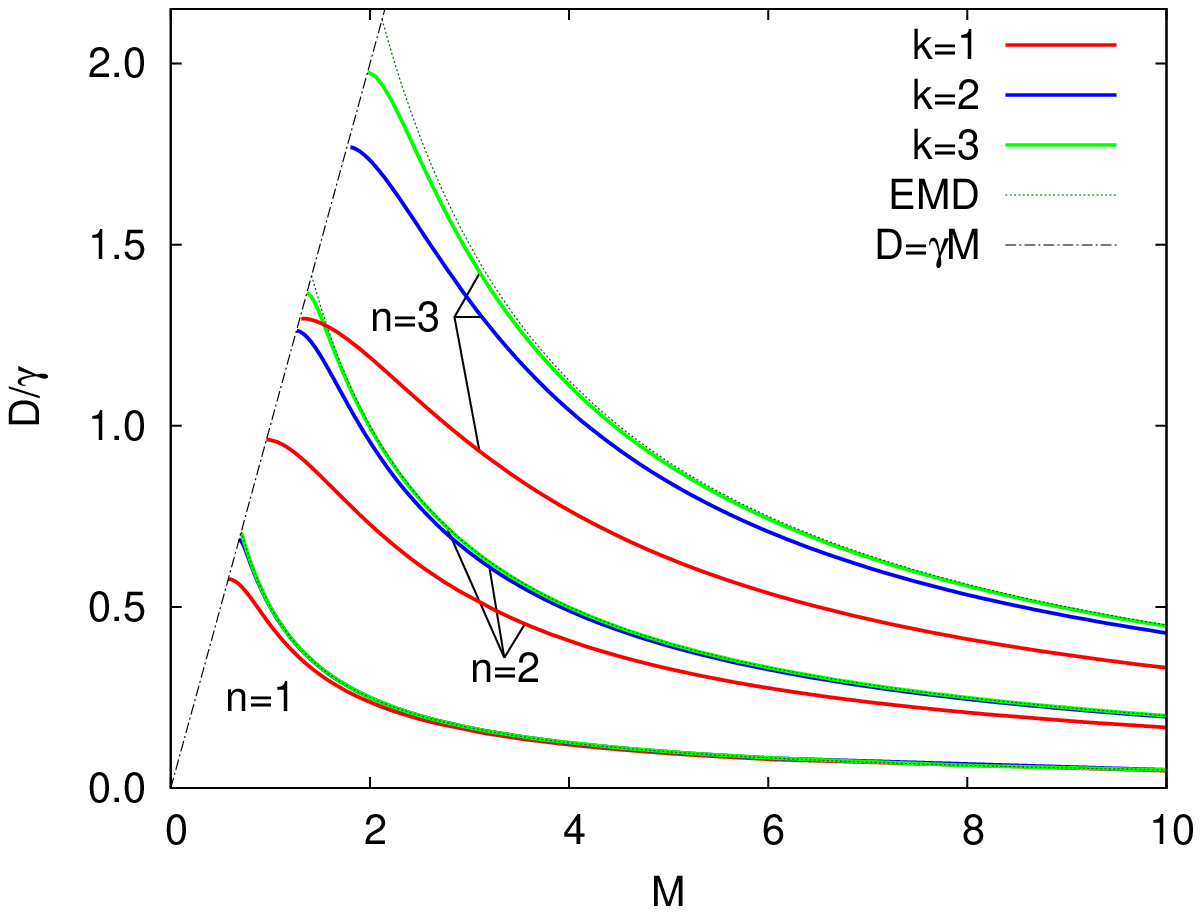}
\hspace*{-0.5cm}
\includegraphics[height=.245\textheight, angle =0]{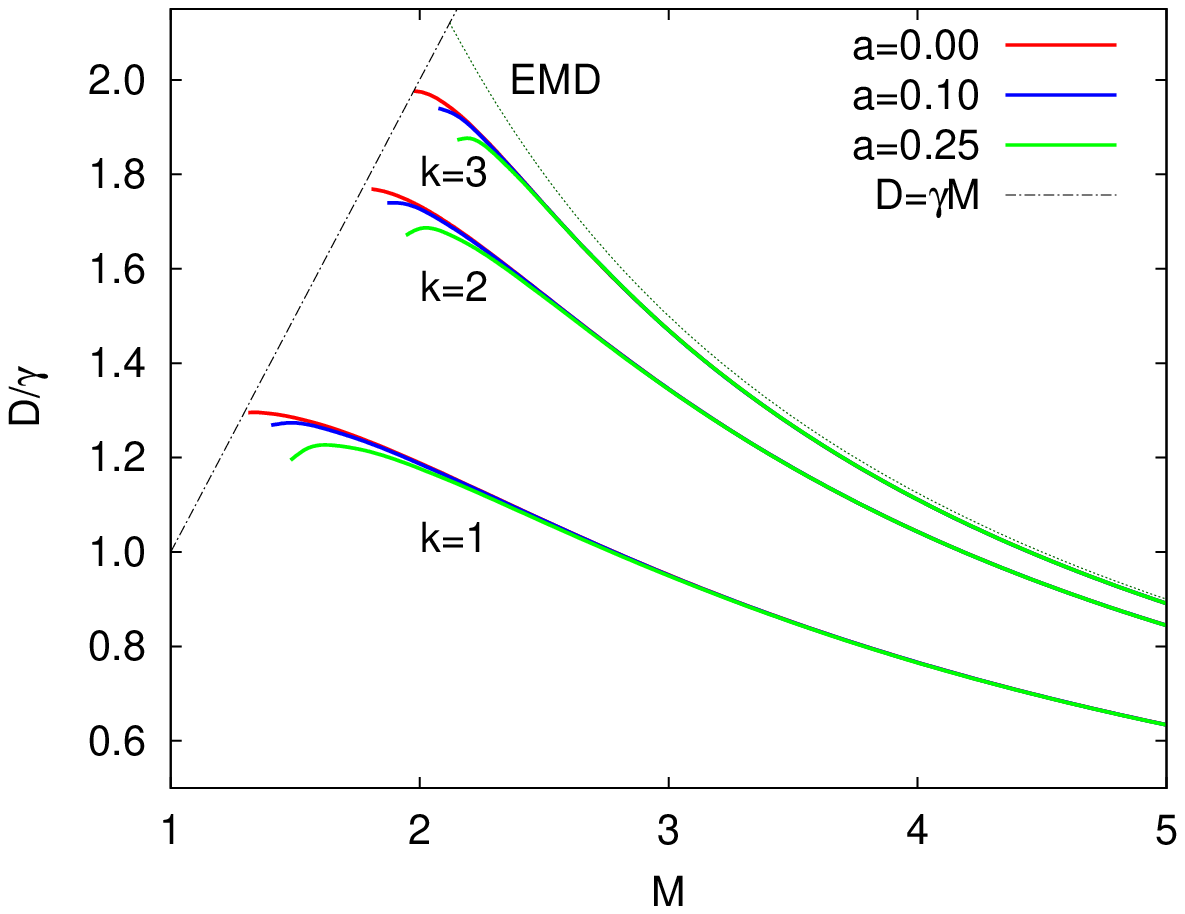}
}
\vspace{-0.5cm}
\caption{
EYMD black holes are compared to 
EMD black holes (thin dotted) with $Q=0$ and $P=n$:
The relative dilaton charge $D/\gamma$ is shown as a function
of the mass $M$ ($\gamma=1$).
a) Static black holes with $n=1-3$ and $k=1-3$.
b) Black holes with specific angular momentum $a=0.25$, 0.1, 0,
$n=3$ and $k=1-3$
} 
\label{fig5}
\end{figure}

The global charges and the horizon properties
of the non-Abelian black holes with node number $k=1$
are already rather close to those of the embedded EMD solutions
with $Q=0$ and $P=n$.
Moreover, also the EYMD metric and dilaton functions as well as 
their gauge field functions (which do not vanish in the static limit)
are impressively close to their
EMD counterparts \cite{Kleihaus:2003sh}.
With increasing node number $k$, however, the non-Abelian black holes
get increasingly closer to the Abelian black holes, 
converging pointwise in the limit $k \rightarrow \infty$.
This convergence can be seen in Fig.~5, where the relative dilaton charge
is exhibited for EYMD solutions with $n=1-3$ and $k=1-3$
together with the corresponding limiting EMD black holes.

Let us finally reconsider the uniqueness conjecture for EYMD black holes.
One could try to replace the magnetic charge $P$ 
by the dilaton charge $D$ or by $D/\gamma$,
since such a replacement occurs in the non-Abelian mass
formula, Eq.~(\ref{nasmarr}).
However, to reinstate an albeit modified uniqueness conjecture, such a 
replacement is not sufficient as seen already in the static case,
where EYMD black holes carry only mass $M$ and dilaton charge $D$,
while $J=Q=0$, as demonstrated in Fig.~5a
for black holes with $n=1-3$, $m=1$, and $k=1-3$ and $\gamma=1$.
As seen in the figure, black holes with the same winding number $n$ 
and different node number $k$ do not intersect, whereas
black holes with different winding numbers $n$ can intersect.
The $n=3$, $k=1$ black holes, for instance, intersect the
$n=2$ black holes for all $k>2$.

To exploit this observation one can introduce a topological charge $N=n$,
putting $N=0$ for embedded Abelian black holes,
and consider a new uniqueness conjecture:
{\sl black holes in SU(2) EYMD theory
are uniquely determined by their mass $M$, their angular momentum $J$,
their non-Abelian electric charge $Q$, their dilaton charge $D$,
and their topological charge $N$}
\cite{Kleihaus:2003sh}.
This uniqueness conjecture 
is illustrated in Fig.~5b, where the relative dilaton charge
is shown versus the mass for
several values of the specific angular momentum $a$
for black holes with $n=3$ and $k=1-3$ ($\gamma=1$).
Also shown are the corresponding limiting Abelian solutions with $P=n$.

\section{Rotating Einstein-Yang-Mills-Higgs-dilaton black holes}

Let us now review the current status of rotating non-Abelian black holes of 
Einstein-Yang-Mills-Higgs-dilaton (EYMHD) theory,
where a Higgs triplet is included in the action.
While we will first present the full set of equations, we will subsequently
focus on the formulae associated with the Higgs field, 
since we can retain most of the discussion for the metric, the gauge fields
and the dilaton from the previous section.
We will then highlight the new features due to the presence of the Higgs field.

\subsection{Action}

We now consider the SU(2) EYMHD action with
matter Lagrangian ${L}_M$ 
\begin{eqnarray}
{L}_M=&&-\frac{1}{2}\partial_\mu \Psi \partial^\mu \Psi
 - \frac{1}{2} e^{2 \gamma \Psi } {\rm Tr} (F_{\mu\nu} F^{\mu\nu})
\nonumber \\
&&-\frac{1}{4} {\rm Tr} \left( D_\mu \Phi D^\mu \Phi \right)
-\frac{\lambda}{8} e^{-2 \gamma \Psi } {\rm Tr} 
 \left( \Phi^2 - v^2 \right)^2
\ , \label{lagmh} \end{eqnarray}
where a 
Higgs field in the adjoint representation $\Phi = \tau^a \Phi^a$
has been coupled to the gauge fields and the dilaton field,
i.e.,
$D_\mu = \nabla_\mu + i e \left[A_\mu , \cdot \ \right] $
represents the gauge covariant derivative,
$\lambda$ is the 
Higgs self-coupling constant, and $v$ denotes the Higgs vacuum
expectation value. The non-Abelian SU(2) symmetry is broken
to an Abelian U(1) symmetry by the nonzero vacuum expectation value of the Higgs field.
The theory then contains a massless photon,
two massive vector bosons of mass $M_W = e v$,
and a massive Higgs field $M_H = {\sqrt {2 \lambda}}\, v$.
In the limit $\lambda = 0$, the Higgs field becomes massless as well.

Variation of the action with respect to the metric and the matter fields
leads, respectively, to the Einstein equations 
with stress-energy tensor
\begin{eqnarray}
T_{\mu\nu} 
  &=& \partial_\mu \Psi \partial_\nu \Psi
     -\frac{1}{2} g_{\mu\nu} \partial_\alpha \Psi \partial^\alpha \Psi
      + 2 e^{2 \gamma \Psi }{\rm Tr}
    ( F_{\mu\alpha} F_{\nu\beta} g^{\alpha\beta}
   -\frac{1}{4} g_{\mu\nu} F_{\alpha\beta} F^{\alpha\beta})
\nonumber\\
&+&
\frac{1}{2} {\rm Tr} \left( D_\mu \Phi D_\nu \Phi - \frac{1}{2} g_{\mu\nu}
 D_\alpha \Phi D^\alpha \Phi \right)
- \frac{\lambda}{8} g_{\mu\nu}  e^{-2 \gamma \Psi }{\rm Tr}
 \left( \Phi^2 - v^2 \right)^2
\ , \label{tmunuh}
\end{eqnarray}
and the matter field equations,
\begin{eqnarray}
& & D_\mu(e^{2 \gamma \Psi } F^{\mu\nu}) =
  \frac{1}{4} i e \left[\, \Phi, D^\nu \Phi \, \right]  \ ,
\label{feqAh} \end{eqnarray}
\begin{eqnarray}
& &\Box \Psi =
 \gamma e^{2 \gamma \Psi }
  {\rm Tr} \left( F_{\mu\nu} F^{\mu\nu} \right)  
 -\frac{\lambda}{4} \gamma e^{-2 \gamma \Psi } {\rm Tr}
 \left( \Phi^2 - v^2 \right)^2
\ ,
\label{feqDh} \end{eqnarray}
where $\Box \Psi = \Psi_{;\sigma}^{\ \ ;\sigma}$, and
\begin{eqnarray}
& &D_\mu D^\mu \Phi = \lambda e^{-2 \gamma \Psi } {\rm Tr}
 \left( \Phi^2 - v^2 \right)  \Phi \ .
\label{feqHh} \end{eqnarray}

\subsection{Higgs field Ansatz and boundary conditions}

Looking for stationary rotating solutions, the symmetry requirements
on the metric, the dilaton field and the non-Abelian gauge field as discussed in
section 2.2 are supplemented with the analogous requirements on the Higgs field
\cite{Forgacs:1979zs}
\begin{equation}
{\displaystyle {\cal L}_{\xi} \Phi = ie[ \Phi, W_{\xi}] \ , \ \ \
 {\cal L}_{\eta} \Phi = ie[ \Phi, W_{\eta} ] }
\ .  \label{symh} \end{equation}
Likewise, the previous Ans\"atze for the 
metric, the dilaton field and the non-Abelian gauge field
are complemented by an
appropriate Ansatz for the Higgs field, given by 
\cite{Kleihaus:2004gm,Kleihaus:2005fs,Kleihaus:2007vf}
\begin{equation}
\Phi =v \left( \Phi_1 \tau_r^{(n,m)} + \Phi_2 \tau_\theta^{(n,m)} \right)
\ , \label{a4} \end{equation}
where $\Phi_1$ and $\Phi_2$ are functions of the coordinates $r$ and $\theta$ only.

The boundary conditions for the metric and the dilaton field
are given in section 2.3.

The boundary conditions for the gauge field and the Higgs field 
in the asymptotic region can again be obtained by a
gauge transformation of some gauge potential $A^{\infty}$ and Higgs field $\Phi^{\infty}$ with 
gauge transformation matrix of the form Eq.~(\ref{gauge}).
It depends on the number of nodes $k$ and the integer $m$.

For odd  $m$ the solutions possess a magnetic charge. In this case
\begin{equation}
A^{\infty} = \nu\frac{\tau_r^{(n,1)}}{2}dt+
\frac{\tau_\vphi^{(n)}}{2} d\theta
-n \sin\theta \frac{\tau_\theta^{(n,1)}}{2} d\varphi \ , \ \ \ 
\Phi^{\infty} = v \tau_r^{(n,1)} , 
\label{rWuYang}
\end{equation} 
and $\Gamma = (1-m)\theta$ for $k$ even, resp. $\Gamma = 0$ for $k$ odd.
This yields the boundary conditions at $r=\infty$ for even $k$
\begin{equation}
\Phi_1=v \ , \Phi_2=0 \ ,\ B_1=\nu \ , \ B_2=0 \ , \ 
\label{bcinfH1}
\end{equation} 
\begin{equation}
H_1=0\ , \ 1-H_2 = m \ , \ 
H_3 = \frac{\cos\theta -\cos(m\theta)}{\sin\theta}\ , \ 
1-H_4 = \frac{\sin(m\theta)}{\sin\theta}  \ , \ 
\label{bcinfH2}
\end{equation} 
and for odd $k$
\begin{equation} 
\Phi_1 = \cos((1-m)\theta)\ , \ 
\Phi_2 = \sin((1-m)\theta)\ ,  
\end{equation}
\begin{equation} 
B_1 = \nu\cos((1-m)\theta)\ , \ 
B_2 = \nu\sin((1-m)\theta)\ ,  
\end{equation}
\begin{equation} 
H_1 =0 \ , \ 1-H_2=1 \ , 
H_3 = -\sin((1-m)\theta)\ , \ \ \ 1-H_4 =\cos((1-m)\theta)\ .
\end{equation}

On the other hand for even $m$ the magnetic charge vanishes. In this case
\begin{equation}
A^{\infty} = \nu\frac{\tau_z}{2}dt \ , \ \ \ 
\Phi^{\infty} = v \tau_z , 
\label{aWuYang}
\end{equation} 
and $\Gamma = -m\theta$ for $k$ odd, resp. $\Gamma = 0$ for $k$ even.
In case $k$ is odd, the resulting boundary conditions are the same as in 
Eqs.~(\ref{bcinfH1}) and (\ref{bcinfH2}), except that now 
\begin{equation} 
H_3 = \frac{1 -\cos(m\theta)}{\sin\theta}\ . \ 
\end{equation}
For even $k$, we find
\begin{equation} 
\Phi_1 = \cos(m\theta)\ , \ \ \ 
\Phi_2 = -\sin(m\theta)\ ,  
\end{equation}
\begin{equation} 
B_1 = \nu\cos(m\theta)\ , \ \ \ 
B_2 = -\nu\sin(m\theta)\ ,  
\end{equation}
\begin{equation} 
H_1 = H_3 =0 \ , \ \ \ 
1-H_2 = 1- H_4 =0 \ .
\end{equation}

The boundary conditions at the horizon and the symmetry axis are the
same as for EYMD black holes.
The Higgs field functions satisfy
\begin{equation}
\partial_r \Phi_1|_{r=r_{\rm H}}= \partial_r \Phi_2|_{r=r_{\rm H}}= 0
\  \label{bh2c} \end{equation}
at the horizon.
The boundary conditions along the symmetry axis
($\theta=0$ and $\theta=\pi$)  are given by
\begin{eqnarray}
    \Phi_2|_{\theta=0}=0 \ , \ \ \ \partial_\theta \Phi_1|_{\theta=0}=0 \ ,
\label{bc4e} \end{eqnarray}
where the analogous conditions hold for $\theta=\pi$.

We again introduce dimensionless quantities,
defining the dimensionless constants $\alpha$, $\beta$, and $\gamma$
\begin{equation}
\alpha = \sqrt{4\pi G} v \ , \ \ \ 
\beta = \sqrt{\frac{\lambda}{e^2}} \ , \ \ \
\bar \gamma =\frac{\alpha}{\sqrt{4\pi G}} \gamma ,
\end{equation}
where $\alpha$ represents the strength of the gravitational
interaction and $\beta$ fixes the ratio of the 
Higgs and vector boson masses,
and the dimensionless coordinate $\bar r$,
\begin{equation}
\bar r=\frac{e \alpha}{\sqrt{4\pi G}} r
\ , \label{dimlessh} \end{equation}
the dimensionless electric gauge field functions
${\bar B}_1$ and ${\bar B}_2$,
\begin{equation}
{\bar B}_1 = \frac{\sqrt{4 \pi G}}{e \alpha}  B_1 \ , \ \ \
{\bar B}_2 = \frac{\sqrt{4 \pi G}}{e \alpha}  B_2 \ ,
\label{barbh} \end{equation}
(i.e., $\bar \nu = \frac{\sqrt{4 \pi G}}{e \alpha} \nu$,)
and the dimensionless dilaton function $\bar \Psi$,
\begin{equation}
\bar \Psi = \frac{\sqrt{4\pi G}}{\alpha} \Psi
\ , \label{dimph} \end{equation}
and subsequently omit the bar for notational simplicity.

\subsection{Global charges}

\noindent{\sl Global charges:}
The expressions for the
mass $M$, the angular momentum $J$ and the dilaton charge $D$
given in section 2.4 remain valid
\cite{Kleihaus:2007vf}.
However, the non-Abelian black holes can now carry Abelian
electric and magnetic charges.
A gauge-invariant definition of
the electromagnetic field strength tensor is
given by the 't Hooft tensor \cite{'tHooft:1974qc}
\begin{equation}
{\cal F}_{\mu\nu} = {\rm Tr} \left\{ \hat \Phi F_{\mu\nu}
- \frac{i}{2e} \hat \Phi D_\mu \hat \Phi D_\nu \hat \Phi \right\}
= \hat \Phi^a F_{\mu\nu}^a + \frac{1}{e} \epsilon_{abc}
\hat \Phi^a D_\mu \hat \Phi^b D_\nu \hat \Phi^c
\ , \label{Hooft_tensor} \end{equation}
where $\hat \Phi$ is the normalized Higgs field,
$|\hat \Phi|^2 = (1/2) {\rm Tr\,} \hat \Phi^2 =\sum_a (\hat \Phi^a)^2 = 1$.

The 't Hooft tensor yields the electric current $j_{\rm el}^\nu$
\begin{equation}
 \nabla_\mu {\cal F}^{\mu\nu} = -4 \pi j_{\rm el}^\nu
\ , \label{jel} \end{equation}
and the magnetic current  $j_{\rm mag}^\nu$
\begin{equation}
 \nabla_\mu {^*}{\cal F}^{\mu\nu} = 4 \pi j_{\rm m}^\nu
\ , \label{jmag} \end{equation}
where ${^*}{\cal F}$ represents the dual field strength tensor.
The electric charge $Q$ is then given by
\begin{equation}
 {Q} = \frac{e}{4\pi} \int_{S_2}
  {^*}{\cal F}_{\theta\varphi}
 d\theta d\varphi 
\ , \label{Qel} \end{equation}
with the integral evaluated at spatial infinity,
and can be read off from the expansion for $B_1$
\begin{equation}
B_1 = \nu -\frac{Q}{r} + O\left(\frac{1}{r^2}\right) .
\label{QP1}
\end{equation}

By rewriting the  't Hooft tensor as
\begin{equation}
{\cal F}_{\mu\nu} = \partial_\mu{\cal A}_\nu - \partial_\nu{\cal A}_\mu
           -\frac{i}{2 e}
 {\rm Tr}\left\{\hat \Phi \partial_\mu \hat \Phi \partial_\nu \hat \Phi \right\}
\ , \label{Hooft_tensor2} 
\end{equation}
with ${\cal A}_\mu = {\rm Tr}\left\{\hat \Phi A_\mu\right\}$,
it follows from Eq.~(\ref{jmag}) that the magnetic current
$j_{\rm m}^\sigma$ is proportional to the topological current $k^\sigma$
\begin{equation}
j_{\rm m}^\sigma = \frac{i}{16\pi e}
\epsilon^{\sigma\rho\mu\nu}
{\rm Tr}\left\{
\partial_\rho\hat \Phi \partial_\mu \hat \Phi \partial_\nu \hat \Phi 
\right\} = \frac{1}{e} k^\sigma
\ . \label{jmag2} 
\end{equation}
Integration of the charge density then leads to the dimensionless
magnetic charge \cite{Kleihaus:2007vf}
\begin{equation}
{P} = {n}\varepsilon  =
\frac{n}{2}\left(1 - (-1)^m\right)
 , \label{QP2} \end{equation}
i.e., $P=n$ for odd $m$, and $P=0$ for even $m$.

\noindent{\sl Physical interpretation of $\nu$:}
The asymptotic expansion of the non-Abelian gauge potential component $A_0$
contains the quantity $\nu$, Eq.~(\ref{QP1}), which is thus not defined in a 
gauge-invariant way. In order to achieve a physical interpretation of $\nu$ one can
apply a gauge transformation given by
$$
U = e^{i \nu t \tau_z/2} \ e^{ i m \theta \tau_\varphi^{(n)}/2}  ,
$$
which leads to an asymptotically trivial 
gauge potential (for even $m$).
Then the transformed gauge potential reads
\begin{eqnarray}
A_\mu dx^\mu
 & = &  \left( \left[\bar B_1 -\nu-n\frac{\omega}{x}(\cos(m\theta)-1)\right]\frac{\tau_z}{2e} 
  +\left[ B_2 +n\frac{\omega}{x}\sin(m\theta)\right]\frac{\tau_\rho^{(n,\nu t)}}{2e}
 \right) dt 
\nonumber \\
& & 
+A_\varphi (d\varphi-\frac{\omega}{x} dt)
+\left(\frac{H_1}{x}dx +(1-H_2-m)d\theta \right)\frac{\tau_\varphi^{(n,\nu t)}}{2e}
 , \label{gt_a1} 
 \end{eqnarray}
with
\begin{equation}
A_\varphi=   -n\sin\theta\left(\left[H_3 +\frac{\cos(m\theta)-1}{\sin\theta}\right]
\frac{\tau_z}{2e}
   +\left[1-H_4-\frac{\sin(m\theta)}{\sin\theta}\right] \frac{\tau_\rho^{(n,\nu t)}}{2e}\right)  
 , \label{gt_a2} \end{equation}
and the transformed Higgs field reads
\begin{equation}
\Phi = \left( \Phi_1 \tau_z + \Phi_2 \tau_\rho^{(n,\nu t)} \right)
\ , \label{gt_a4} \end{equation}
with the explicitly time dependent matrices \cite{Kleihaus:2007vf}
\begin{eqnarray}
\tau_\rho^{(n,\nu t)}  & = & \cos(n \varphi -\nu t) \tau_x + \sin(n \varphi -\nu t) \tau_y \ , 
\nonumber\\
\tau_\varphi^{(n,\nu t)} & = & -\sin(n \varphi -\nu t) \tau_x + \cos(n \varphi -\nu t) \tau_y \ .
\nonumber
\end{eqnarray}
Consequently, in this gauge the fields are explicitly time dependent and rotate 
in the internal space about the $\tau_z$ direction, where the quantity $\nu$ corresponds to 
the rotational frequency \cite{Kleihaus:2007vf}.
We note, that in the presence of a magnetic charge, i.e. for odd $m$, the transformed gauge potential 
is singular on the negative $z$ axis, but the physical interpretation of $\nu$ does not
change.
We further note, that the phase $(n \varphi -\nu t)$ 
precisely corresponds to the phase of the scalar field
encountered in rotating boson stars
\cite{Schunck:1996he}.

\noindent{\sl Mass formula:}
As shown in \cite{Kleihaus:2007vf},
the stationary axially symmetric EYMHD black hole solutions
satisfy the mass formula
\begin{equation}
M = 2 {T}{S} + 2 \Omega J 
 + 2{\Phi}_{\rm el} Q (1-\varepsilon)  +2\nu Q 
+ \frac{D}{\gamma}
, \label{namass2}
\end{equation}
where 
$\varepsilon=\frac{1}{2}\left(1 - (-1)^m\right)$.
Interestingly, 
this mass formula differs from the mass formula (\ref{nasmarr}),
valid for EMD and EYMD black holes.
First of all,
the third term is only present for magnetically neutral
black holes, where $\varepsilon=0$, since
for magnetically charged black holes ($\varepsilon=1$) it vanishes.
Moreover, the fourth term is an additional term which is
not present for EMD and EYMD black holes. 
This term contributes to the mass of all
electrically charged EYMHD black holes.
Intriguingly, it contains the electric gauge potential parameter $\nu$,
which was interpreted as a rotational frequency. 

Let us now compare this mass formula with other known mass formulae.
First we consider globally regular EYMHD solutions.
Here the mass formula simplifies to
$M=2\nu Q + \frac{D}{\gamma}$ \cite{Kleihaus:2007vf}.
Indeed, when the black hole horizon size is shrunk to zero size, 
the first term vanishes, while the second and the third term cancel each other.

Next let us consider the recently discovered 
rotating black holes with scalar hair \cite{Herdeiro:2014goa}.
For these the mass formula reads
\begin{equation}
M = 2 {T}{S} + 2 \Omega J 
      - 2 \nu Q + M^\Phi ,
\end{equation}
where $M^\Phi$ corresponds to the mass contribution from the scalar field
outside the horizon. By including a dilaton this term could be covered by 
$D/\gamma$. More interesting is the third term, which we have here
expressed in terms of the rotational frequency of the scalar field $\nu$,
making use of the relation
$\nu =  \Omega n$ \cite{Herdeiro:2014goa}.
In this form this term is seen to formally agree 
with the one in the above EYMHD mass formula (up to a sign).
We note, however, that for EYMHD black holes, $\nu$ and $\Omega$ are
independent.

\subsection{Nonperturbative black holes}

The numerical EYMHD black holes are obtained analogously
to the EYMD black holes.
For given coupling constants $\alpha$, $\beta$ and $\gamma$,
the rotating non-Abelian black holes
then depend on the horizon radius $r_{\rm H}$,
the rotational velocity of the horizon $\Omega$,
the gauge potential parameter $\nu$
and the integers $m$, $n$ and $k$.

\subsubsection{EYMH black holes}

\begin{figure}[t]
\mbox{
\includegraphics[height=.27\textheight, angle =0]{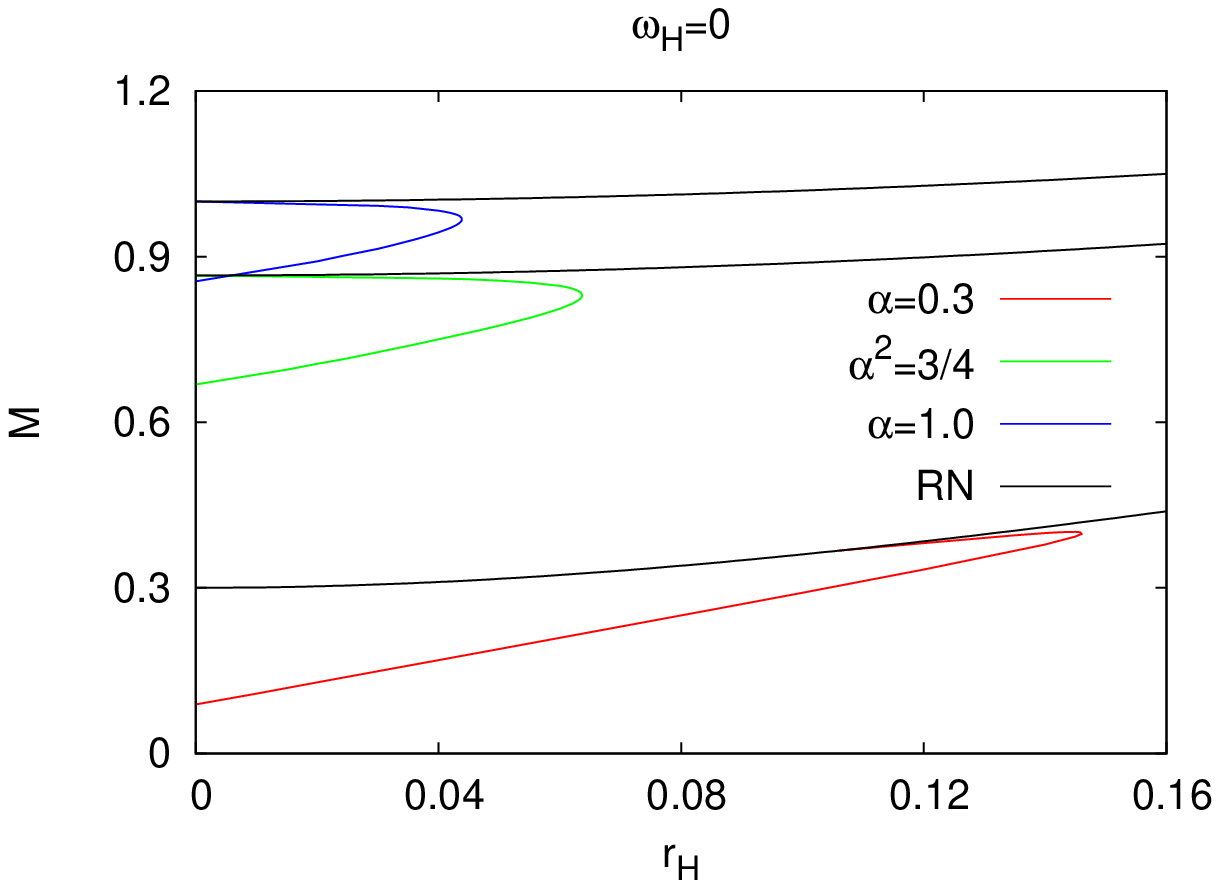}
\hspace*{0.5cm}
\includegraphics[height=.27\textheight, angle =0]{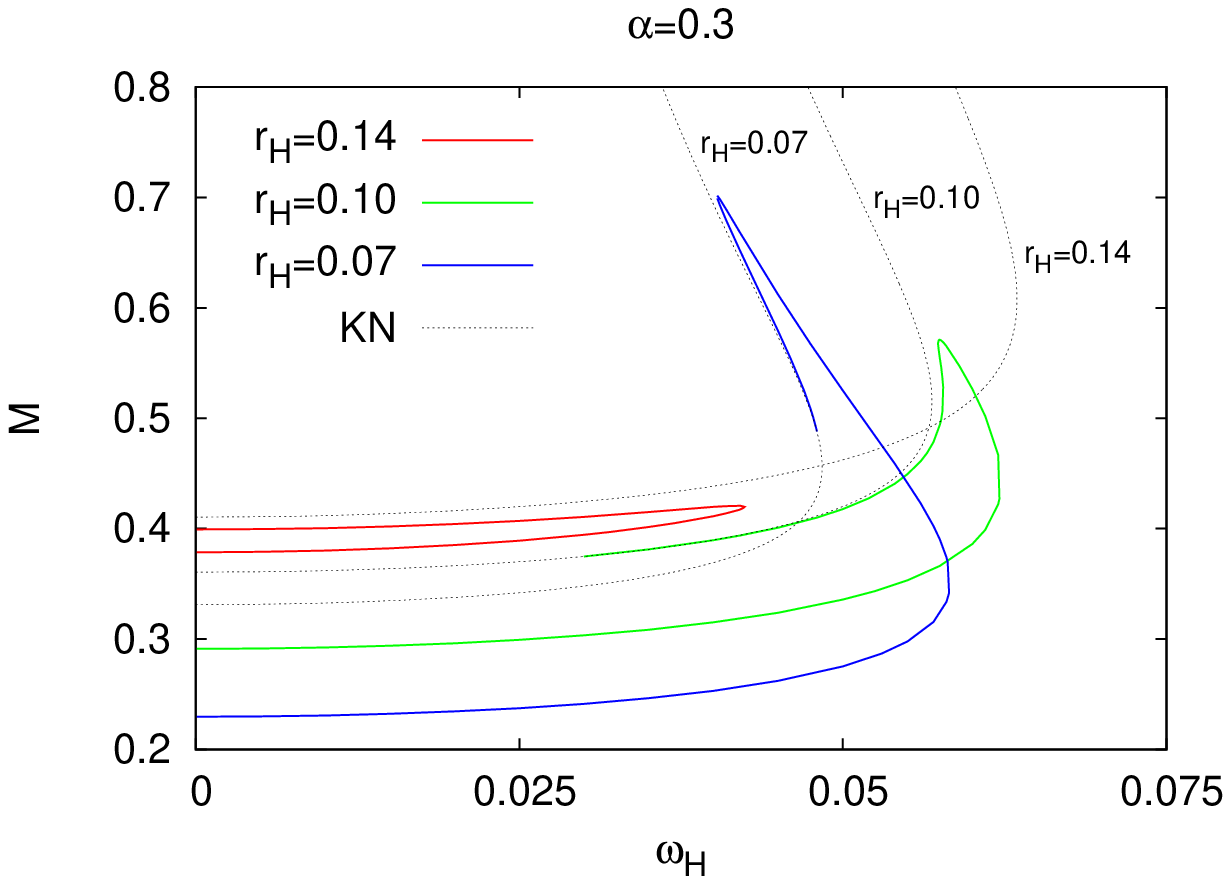}
}
\mbox{
\includegraphics[height=.27\textheight, angle =0]{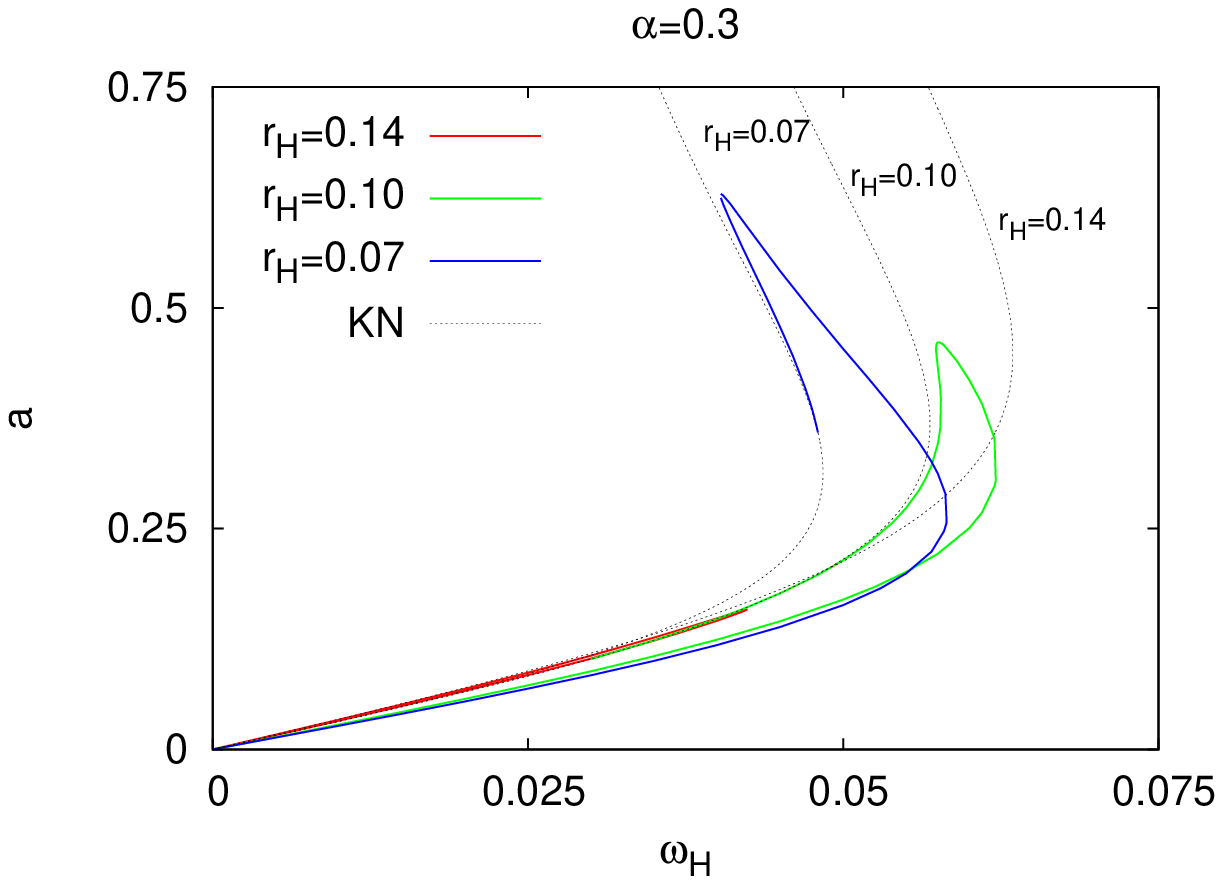}
\hspace*{0.5cm}
\includegraphics[height=.27\textheight, angle =0]{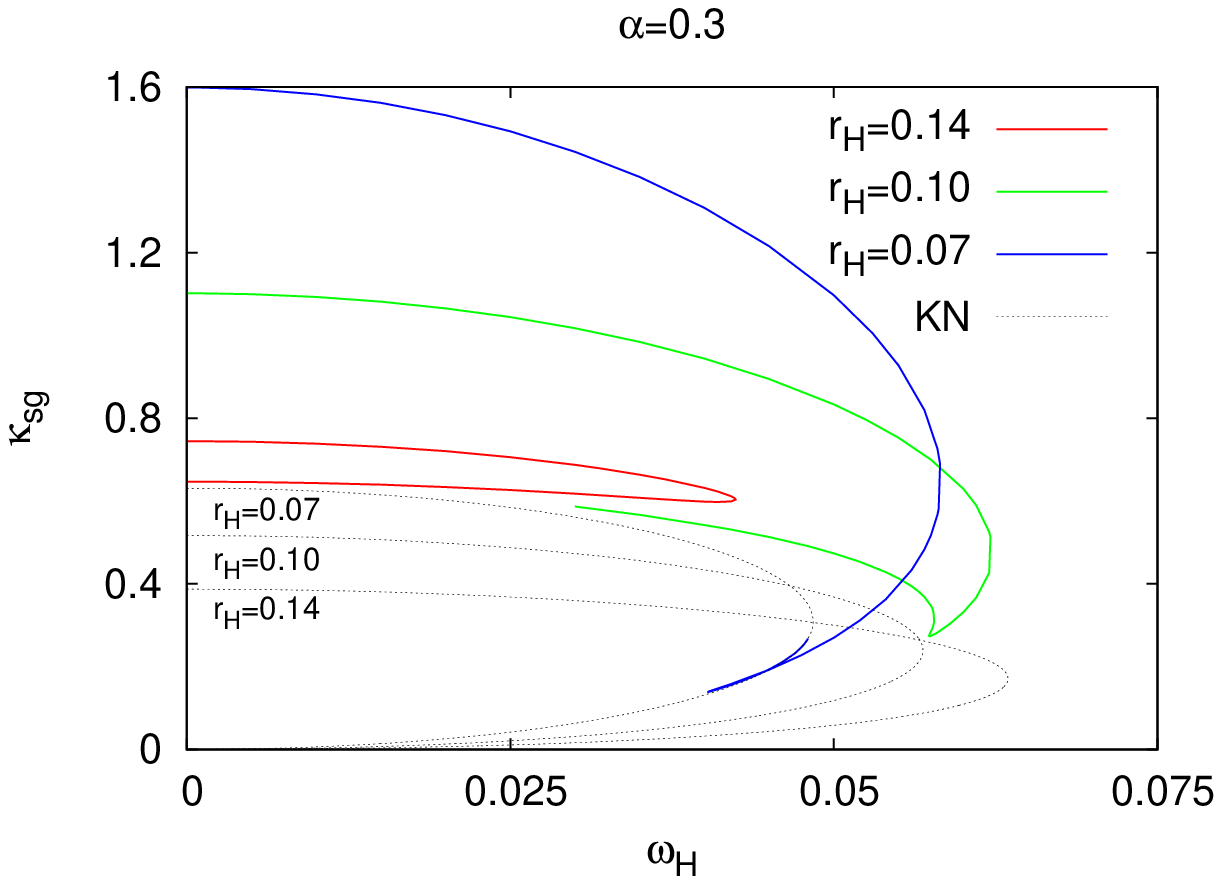}
}
\vspace{-0.5cm}
\caption{
EYMH black holes ($n=1$, $m=1$, $k=0$, $\beta=\gamma=0$, $\nu=0$):
a) The dimensionless mass $M$ is shown as a function of 
the isotropic horizon radius $r_{\rm H}$ for static black holes
and several values of $\alpha$.
Also shown are embedded RN
black holes (dotted) with $Q=0$ and $P=1$.
b) 
The dimensionless mass $M$ is shown as a function of $\omega_{\rm H}$ for
for rotating black holes for several values of
$r_{\rm H}$ and $\alpha=0.3$.
Also shown are embedded KN
black holes (dotted) with $Q=0$ and $P=1$.
c) Same as b) for the specific angular momentum $a=J/M$ .
d) Same as b) for the surface gravity $\kappa_{\rm sg}$.
}
\label{fig6}
\end{figure}

We again first consider the simplest
type of rotating black holes.
These EYHM black holes carry unit magnetic charge 
and no electric charge. They
emerge from the static spherically symmetric fundamental 
black holes with monopole hair ($n=1$, $m=1$, $k=0$),
when a finite horizon angular velocity is imposed.
In the following we discuss their properties and
their observed critical behavior. This includes in particular
their bifurcation with embedded Kerr-Newman black holes. 

To better understand the observed pattern let us briefly recall
the static case, starting with the regular gravitating monopoles.
Gravitating monopoles exist only 
for sufficiently small gravitational coupling, $0  < \alpha < \alpha_{\rm max}$,
where $\alpha_{\rm max}$ is on the order of one
and decreases with increasing $\beta$
\cite{Lee:1991vy,Breitenlohner:1991aa,Breitenlohner:1994di}.
At $\alpha_{\rm max}$ a bifurcation of the 
branch of gravitating monopoles
with the set of embedded extremal RN
black holes with electromagnetic charges $Q=0$, $P=1$ is encountered,
unless $\beta$ vanishes or $\beta$ is very small.
(In that case, a very short second branch of gravitating monopoles
is present for $\alpha_{\rm cr} < \alpha \le \alpha_{\rm max}$, bifurcating at
$\alpha_{\rm cr}$ with the set of extremal RN black holes.)

Static magnetically charged EYMH black hole solutions
emerge from the gravitating monopole solutions
when a finite regular event horizon is imposed,
representing ``black holes within magnetic monopoles''
\cite{Lee:1991vy,Breitenlohner:1991aa,Breitenlohner:1994di}.
Since they are distinct from their embedded RN counterparts,
they carry non-Abelian hair.
An important feature, which distinguishes these black holes
from the EYM black holes discussed in section 2, is that the size 
of their horizon is bounded.
They cannot become arbitrarily large.
The mass of these static EYMH black holes is shown
in Fig.~6a as a function of the isotropic horizon radius $r_{\rm H}$
for several values of the gravitational coupling strength.

For small values of $\alpha$, 
the black holes merge into non-extremal RN black holes
at a critical value of the horizon radius, whereas
for values of $\alpha$ above a certain value
$\alpha_{\rm tr} = \sqrt{3}/2$
the black holes bifurcate with extremal RN solutions
\cite{Lee:1991vy,Breitenlohner:1991aa,Breitenlohner:1994di}.
(Recall, that in isotropic coordinates $r_{\rm H}=0$ corresponds to regular solutions
on the lower branch and extremal solutions on the upper branch.)
The figure also shows, that in a certain range of parameters
the Abelian RN black holes are unstable to grow non-Abelian hair
\cite{Lee:1991vy,Breitenlohner:1991aa,Breitenlohner:1994di}.

Let us now include rotation, keeping all parameters fixed
except for the horizon angular momentum of the black holes,
which is increased from zero \cite{Kleihaus:2004gm}. 
Inspection of the resulting rotating EYMH black holes then reveals 
another interesting feature.
In contrast to the EYM black holes discussed in section 2,
the rotation does not induce an electric charge
for these EYMH black holes.
Of course, the rotation induces an electric component of the gauge field.
However, this only yields a gravitoelectric moment
and no electric charge \cite{Kleihaus:2004gm}. 
In order to obtain dyonic EYMH black holes,
possessing both magnetic and electric charge,
the electric gauge potential parameter $\nu$ must be given a finite value
\cite{Brihaye:1998cm,Kleihaus:2007vf}.

As illustrated in Fig.~6b,
for fixed horizon radius $r_{\rm H}$
a branch of rotating EYMH black holes
emerges from the corresponding (lower mass) static solution
and extends up to a maximal value 
$\omega_{\rm H}^{\rm max}$
and then bends backwards as expected.
The critical behavior then depends on the value of $\alpha$ and
the value of $r_{\rm H}$.
For $\alpha < \sqrt{3}/2$, and 
$r_{\rm H}$ below the corresponding critical value,
the rotating non-Abelian black holes merge with 
a non-extremal KN black hole,
as seen in Fig.~6a for $r_{\rm H}=0.07$ and 0.10.
Clearly, the instability w.r.t.~growing YMH hair, present for RN black holes,
is perpetuated by KN black holes.

As $r_{\rm H}$ is increased further,
it becomes clear from Fig.~6a, that the critical value of the horizon radius
is passed, and
two static non-Abelian solutions exist.
The rotating EYMH black holes then no longer bifurcate
with the KN black holes.
Instead they lead from the lower static
EYMH black hole to the upper static EYMH black hole,
as seen in Fig.~6b for $r_{\rm H}=0.14$.
This pattern holds as well for $\alpha > \sqrt{3}/2$,
independent of the horizon size.
The further properties of these rotating EYMH black holes 
also reflect this pattern, as seen in Fig.~6c,
where we exhibit the specific angular momentum $a$
as a function of $\omega_{\rm H}$,
and in Fig.~6d, where the surface gravity is illustrated.

Let us now briefly address the EYMH black holes
with higher integers $n$, $m$, $k$,
which have been investigated in detail in the static case,
while a systematic study of the properties 
of their rotating counterparts is still missing.
Interestingly, the radial excitations ($k>0$) of the 
static spherically symmetric ($n=m=1$) non-Abelian black holes
are related to the EYM black holes and are unstable 
\cite{Breitenlohner:1991aa,Breitenlohner:1994di,Hollmann:1994fm}.
When set into rotation, they are expected to retain 
(at least some of) their unstable modes.

The static axially symmetric black holes with higher charges $P=n>1$,
i.e., black holes within multimonopoles,
follow the same pattern as the one
observed for black holes with charge $P=n=1$, 
where 
the value $\alpha_{\rm tr} = \sqrt{3}/2$ of the gravitational coupling
is playing the same critical role
\cite{Hartmann:2001ic}.
Therefore, we conjecture that also the axially symmetric rotating black holes
with higher charges follow the pattern of the rotating black holes
with unit magnetic charge.
The fundamental ($k=0$) static and rotating EYMH black holes with 
$P=n=2$ (and $m=1$)
should again be stable in a certain parameter range, 
and the embedded Abelian rotating black holes with $P=2$ should
be unstable in a corresponding parameter range.

For magnetic charges $P=n>2$, however,
besides the axially symmetric configurations also
monopole configurations with crystal symmetries arise
\cite{Ridgway:1994sm,Ridgway:1995ke,Ridgway:1995ac}.
In contrast to the axially symmetric higher charge monopoles, these
gravitating monopoles with discrete symmetries have been obtained only
perturbatively so far. 
But only a non-perturbative approach would allow to 
find out, which of the possible configurations would be the energetically lowest one
and thus stable within a given sector $P=n$.
(Recall that in flat spacetime for vanishing Higgs mass, the
monopole configurations for a given charge $P=n$ (and $m=1$) are all degenerate
in mass.) 
Clearly, independent of their symmetry, all regular monopole solutions should be
generalizable to contain a - sufficiently small - black hole.
However, we expect the existence of stationary
rotating black hole configurations only for the axially symmetric configurations.

Still different types of black holes are obtained, when the integer
$m$ is increased. For small $n$, the 
associated regular solutions then correspond to
monopole-antimonopole pairs ($m=2$) or
monopole-antimonopole chains ($m>2$), 
consisting of a total of $m$ alternating magnetic poles
located on the symmetry axis. Depending on the Higgs mass, 
for the larger values of $n$ also
systems with vortex rings arise,
where the modulus of the Higgs field vanishes on one or more
rings located in the equatorial plane and in parallel planes
\cite{Kleihaus:2000hx,Kleihaus:2004fh,Kunz:2007jw,Kleihaus:2007vf}.
For odd $m$, these configurations carry a total magnetic charge $P=n$, whereas for
even $m$, they are magnetically neutral.

Again, these configurations can be augmented with a small black hole
at their center. In the simplest case black holes with magnetic
dipole hair arise \cite{Kleihaus:2000kv}. Here again only the static
configurations have been studied in detail, while their rotating
generalizations are still waiting to be explored in a systematic way
\cite{Kleihaus:2007vf}.
Finally, all of the above configurations can be endowed with electric charge
as discussed below
\cite{Kleihaus:2007vf}.

\subsubsection{EYMHD black holes}

\begin{figure}\centering
a\mbox{\epsfysize=7.5cm  \epsffile{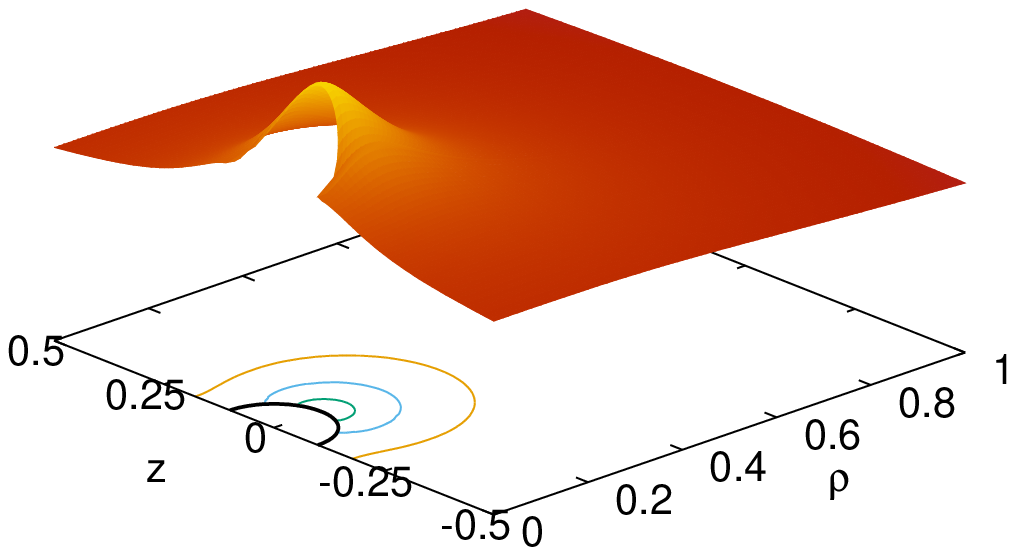}  }
\vspace{1.cm}\\
\begin{tabular}{ccc}
b & c & d \\
\mbox{\hspace*{-15mm}\epsfysize=5cm\epsffile{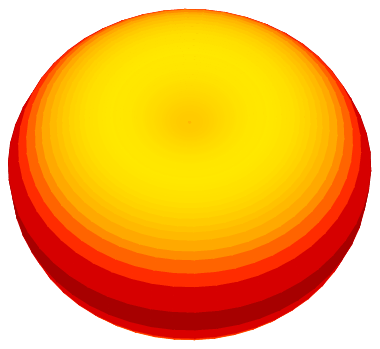}\hspace*{-15mm}}
&
\raisebox{0.mm}{\hspace*{-15mm}\epsfysize=5cm\epsffile{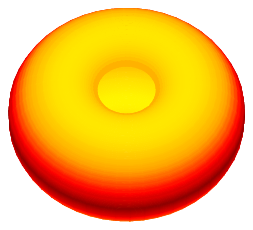}\hspace*{-15mm}}
&
\raisebox{0.mm}{\hspace*{-15mm}\epsfysize=5cm\epsffile{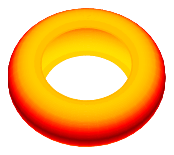}\hspace*{-15mm}}
\\
$\varepsilon = 0.010$ &
$\varepsilon = 0.015$ &
$\varepsilon = 0.020$
\end{tabular}
\caption{
The component $\varepsilon=-T_0^0$ of the stress energy tensor is shown as a
function of the coordinates $\rho = r \sin\theta$, $z=r\cos\theta$
for a dyonic rotating black hole ($n=1$, $m=1$, $k=1$,
$r_{\rm H}=0.1$, $\Omega=0.5$,
$\nu=0.04$, $\alpha=0.3$, $\beta=0.1$, $\gamma=0.1$).
Also shown are surfaces of constant $\varepsilon$.
}
\label{fig7}
\end{figure}

Let us now address EYMHD black holes
by allowing for a finite dilaton coupling $\gamma$.
Considering the dilaton as a kind of scalar graviton, one expects that
the dilaton coupling $\gamma$ is also bounded from above,
just like the gravitational coupling $\alpha$.
As shown in \cite{Brihaye:2001bk}, this is indeed the case,
where the boundary in the parameter space $\gamma_{\rm max}(\alpha)$ 
shrinks with increasing horizon radius $r_{\rm H}$,
excluding the existence of large EYMHD black holes.

So far, the parameter space of rotating EYMHD black holes
has not been explored systematically.
However, a number of calculations have been performed,
suggesting that the basic picture known form EYMH black holes
remains valid, except that the bifurcations occur with rotating
EMD black holes.
Thus EMD black holes should be unstable to growing YM hair as well
in a certain parameter range just like their KN counterparts.

An electric charge can be included, by imposing a
finite value of the electric gauge potential parameter, $\nu \ne 0$.
We exhibit in Fig.~7 an example of such a dyonic rotating black hole,
which has $n=1$, $m=1$, $k=1$, horizon radius $r_{\rm H}=0.1$,
horizon angular velocity $\Omega=0.5$, electric gauge potential
parameter $\nu=0.04$, and was
obtained for the coupling constants $\alpha=0.3$, $\beta=0.1$, and $\gamma=0.1$.
The figure illustrates again the component $\varepsilon=-T_0^0$ of the stress energy tensor.
Thus a black hole within a dyon 
can rotate, whereas a regular dyon
by itself cannot \cite{Heusler:1998ec,VanderBij:2001nm,vanderBij:2002sq}.

Except for spherically symmetric configurations,
the presence of electric and magnetic fields gives rise to an
angular momentum density 
\cite{VanderBij:2001nm,Paturyan:2004ps,Kleihaus:2005fs}.
But globally regular EYMH solutions with a
magnetic charge cannot rotate in the usual sense, since their
angular momentum vanishes,
whereas globally regular EYMH solutions without a magnetic charge
can rotate, i.e., possess a finite angular momentum.
Interestingly, their angular momentum is
proportional to their electric charge 
and thus quantized \cite{VanderBij:2001nm}, i.e.,
analogous to boson stars $J=nQ$ \cite{Schunck:1996he}.
However, as seen for the black hole example in Fig.~7,
black holes with magnetic charge
can rotate.
While a number of electrically charged rotating EYMHD black holes
have already been constructed and shown to
satisfy the non-Abelian mass formula (\ref{namass2}),
these black holes with $n \ge 1$ and $m \ge 1$ 
are yet to be studied systematically.

\section{Conclusions}

In the 1990s the discovery of a plethora of static black holes
with non-Abelian hair thoroughly demolished the long-cherished
belief that the no-hair conjecture would be a rather comprehensive concept,
that should hold for numerous theories with matter fields
and, in particular, for theories with non-Abelian gauge fields,
as encountered in the standard model of particle physics
and in grand unified theories.

In hindsight, this former belief seems surprizing, since 
many localized finite energy solutions were known in these
theories, solitons and sphalerons, which can be coupled
with gravity and endowed with a small black hole
in their interior. Arguments for the existence of such
black holes within monopoles or sphalerons could have
been put forward long ago.
Merely the existence of regular solutions without a flat
spacetime limit, such as the BM solutions of pure EYM theory,
and their associated black holes
was less straightforward to anticipate.
However, it was the discovery of the latter,
which led to a surge of interest and a rapidly growing
body of knowledge in the field, as summarized first in
\cite{Volkov:1998cc}.

Once the existence of static black holes with non-Abelian hair was
established, 
the existence of their rotating generalizations was certainly expected,
though their construction was technically more involved, 
since the presence of rotation does no longer allow for spherical symmetry.
While first attempts to include rotation were based on 
perturbation theory \cite{Volkov:1997qb,Brodbeck:1997ek},
subsequently non-perturbative rotating black holes with non-Abelian hair
could be constructed numerically
\cite{Kleihaus:2000kg,Kleihaus:2002ee,Kleihaus:2002tc,Kleihaus:2003sh,Kleihaus:2007vf,Kleihaus:2004gm}.

In the pure EYM case, the onset of rotation 
of the neutral non-Abelian black holes not only induces a non-Abelian
electric field, as would be the case for rotating RN black holes (with magnetic charge), 
but it induces a non-Abelian electric charge. 
Rotating EYM black holes therefore possess three global charges,
mass $M$, angular momentum $J$, and non-Abelian electric charge $Q$. 
This was predicted by perturbation theory as one of three possible types of solutions.
The two remaining types, however, do not seem to exist at a non-perturbative level
in EYM theory. In contrast, in EYMD theory, the presence of the dilaton
allows for solutions of the second type, 
which possess a finite angular momentum and vanishing non-Abelian charge.
The third type of solutions, which should possess a vanishing angular momentum
while being non-static, does not seem to exist in EYMD theory, either.

Many features of these rotating EYM and EYMD black holes
are analogous to those of their static counterparts. For instance,
there is no upper bound for the size of these black holes,
and the sequences of radially excited rotating black holes
tend (in their outer region) to embedded Abelian black holes. 
Moreover, these rotating non-Abelian black holes should inherit
the instabilities of their static counterparts.
Interestingly,
in EYMD theory a Smarr-type mass formula holds 
for these non-Abelian rotating black holes,
since the dilaton term can comprise the mass contribution
originating from the magnetic fields,
when a magnetic charge itself is missing.

The presence of a Higgs field has a number of significant consequences
for the existence and the properties of the black holes
\cite{Lee:1991vy,Breitenlohner:1991aa,Breitenlohner:1994di,Hartmann:2000gx,Hartmann:2001ic,Kleihaus:2004fh,Kleihaus:2005fs,Kleihaus:2007vf,Kleihaus:2000kv}.
First of all, the horizon size of EYMH black holes is bounded from above.
Considering these configurations as black holes within solitons,
i.e., black holes within regular solutions with a localized finite energy density,
the finite of extent of these solitons seems to limit the size of
the black holes they can contain.
This observation looks rather general, holding not only
for EYMH theory with triple and with doublet Higgs fields,
but also for theories with other types of matter fields, like Einstein-Skyrme theory.

In the case of a triplet Higgs, the EYMH black holes carry 
electromagnetic charge(s). But unlike EYM theory, rotation 
induces only a dipole moment and not
an electric charge of the black holes. In order to have electrically
charged black holes one has to choose a particular boundary condition
for the electric component of the gauge field.
This involves a parameter $\nu$, which can be interpreted
as describing a rotational frequency in internal space.
Interestingly, this parameter enters the Smarr-like mass formula
for these non-Abelian black holes 
(which can be obtained, when a dilaton is included).
Formally this term in the mass formula has a counterpart
in the mass formula of rotating black holes within boson stars.

The presence of the embedded RN and KN black holes
allows the branches of genuinely non-Abelian black holes to merge and end
in such embedded Abelian black holes.
Clearly, this signals the presence of an instability of the Abelian
black holes. Indeed, RN and KN black holes in EYMH theory
become unstable in a certain parameter region with respect
to developing non-Abelian hair.
When genuine non-Abelian black holes
can no longer exist, only embedded Abelian black holes remain.

In principle, a plethora of static and rotating black hole configurations
is possible in these non-Abelian theories. 
Stationary axially symmetric configurations can be
characterized by three integers, $n$, $m$ and $k$.
While $k$ labels the radial excitations,
$m$ denotes the number of `components' of composite solutions,
e.g., a monopole-antimonopole pair has $m=2$,
and $n$ represents the magnetic charge of the components.
All regular configurations can be endowed with a black hole
at their center, and all static black holes can be set into rotation.

An interesting open question here is, whether in EYMH theory also 
genuinely non-Abelian systems
of rotating black holes can exist,
which are completely regular outside their horizons.
While Einstein-Maxwell theory 
does not allow for systems of rotating solutions
(see e.g.~\cite{Neugebauer:2011qb} and references therein),
the presence of non-Abelian fields might result in a 
balance of forces and thus regularity.
Finding such solutions would be another highlight
and important example, showing once again
that established Abelian theorems cannot be extended
to the non-Abelian case.

Last but not least let us briefly address the generalization of
such non-Abelian solutions in the presence of a cosmological constant,
yielding Anti-de Sitter (AdS) or de Sitter (dS) black holes.
EYM solutions in asymptotically AdS space exhibit
new interesting features. In particular, they need a further parameter
for their characterization, and
there are stable spherically symmetric black hole solutions,
not present in asymptotically flat space
\cite{Winstanley:1998sn,Bjoraker:1999yd,Bjoraker:2000qd}.
Topological EYM black holes which possess a
nonspherical event horizon topology were also shown to exist in EYM theory
with a negative cosmological constant
\cite{VanderBij:2001ia}.
While static AdS EYM black holes with axial symmetry only have been obtained recently
\cite{Kichakova:2015nni},
rotating generalizations of AdS EYM black holes are still missing.
Only globally regular rotating configurations have been obtained so far
\cite{Radu:2002rv}.
The inclusion of a positive cosmological constant leads to black hole solutions
which possess a cosmological horizon.
However, static spherically symmetric dS EYM solutions 
were shown to be unstable \cite{Torii:1995wv,Brodbeck:1996xm}.
(See the review \cite{Winstanley:2008ac} for further details
on AdS and dS EYM black holes.)

\subsection*{Acknowledgments}
We gratefully acknowledge support by the DFG Research Training Group 1620 ``Models of Gravity'' and by the grant FP7, Marie Curie
Actions, People, International Research Staff Exchange Scheme (IRSES-606096).

\end{document}